\documentclass[aps,eqsecnum,preprint,floats,epsf,epsfig,nofootinbib]{revtex4}
\usepackage{graphicx}
\usepackage{epsfig}

\begin{document}
\def\be{\begin{eqnarray}}
\def\en{\end{eqnarray}}
\def\non{\nonumber}
\def\la{\langle}
\def\ra{\rangle}
\def\nc{N_c^{\rm eff}}
\def\vp{\varepsilon}
\def\drho{\bar\rho}
\def\deta{\bar\eta}
\def\a{{\cal A}}
\def\B{{\cal B}}
\def\c{{\cal C}}
\def\d{{\cal D}}
\def\e{{\cal E}}
\def\p{{\cal P}}
\def\t{{\cal T}}
\def\up{\uparrow}
\def\dw{\downarrow}
\def\vma{{_{V-A}}}
\def\vpa{{_{V+A}}}
\def\smp{{_{S-P}}}
\def\spp{{_{S+P}}}
\def\J{{J/\psi}}
\def\ov{\overline}
\def\Lqcd{{\Lambda_{\rm QCD}}}
\def\pr{{Phys. Rev.}~}
\def\prl{{Phys. Rev. Lett.}~}
\def\pl{{Phys. Lett.}~}
\def\np{{Nucl. Phys.}~}
\def\zp{{Z. Phys.}~}
\def\lsim{ {\ \lower-1.2pt\vbox{\hbox{\rlap{$<$}\lower5pt\vbox{\hbox{$\sim$}
}}}\ } }
\def\gsim{ {\ \lower-1.2pt\vbox{\hbox{\rlap{$>$}\lower5pt\vbox{\hbox{$\sim$}
}}}\ } }

\font\el=cmbx10 scaled \magstep2{\obeylines\hfill February, 2005}

\vskip 1.5 cm

\centerline{\large\bf $B\to f_0(980)K$ Decays and Subleading
Corrections}
\bigskip
\centerline{\bf Hai-Yang Cheng$^{1}$ and Kwei-Chou Yang$^{2}$}
\medskip
\centerline{$^1$ Institute of Physics, Academia Sinica}
\centerline{Taipei, Taiwan 115, Republic of China}
\medskip
\centerline{$^2$ Department of Physics, Chung Yuan Christian
University} \centerline{Chung-Li, Taiwan 320, Republic of China}
\bigskip
\bigskip
\bigskip
\bigskip
\centerline{\bf Abstract}
\bigskip
\small The decay $B\to f_0(980)K$ is studied within the framework
of QCD factorization and the two-quark scenario for $f_0(980)$.
There are two distinct penguin contributions and their
interference depends on the unknown mixing angle $\theta$ of
strange and nonstrange quark contents of $f_0(980)$: destructive
for $0<\theta<\pi/2$ and constructive for $\pi/2<\theta<\pi$.  The
QCD sum rule method is applied to evaluate the leading-twist
light-cone distribution amplitudes and the scalar decay constant
of $f_0$. We conclude that the short-distance approach is not
adequate to explain the observed large rates of $f_0K^-$ and
$f_0\ov K^0$. Among many possible subleading corrections, we study
and estimate the contributions from the three-parton Fock states
of the $f_0$ and from the intrinsic gluon inside the $B$ meson. It
is found that the spectator gluon of the $B$ meson may play an
eminent role for the enhancement of $f_0(980)K$. We point out that
if $f_0(980)$ is a four-quark state as widely perceived, there
will exist extra diagrams contributing to $B\to f_0(980)K$.
However, in practice it is difficult to make quantitative
predictions based on the four-quark picture for $f_0(980)$ as it
involves additional nonfactorizable contributions that are
difficult to estimate and the calculations of the decay constant
and form factors of $f_0(980)$ are beyond the conventional quark
model.

\pagebreak

\section{Introduction}
The decay of the $B$ meson into a scalar meson $f_0(980)$ was
first measured by Belle \cite{Belle1} in the charged $B$ decays to
$K^\pm\pi^\mp\pi^\pm$ and a large branching fraction product for
the $f_0(980)K^\pm$ final states was found. A recent updated
result by Belle yields \cite{Belle2}
 \be \label{eq:Bellef0K1}
 \B(B^+\to f_0(980)K^+\to \pi^+\pi^-K^+)=
 (7.55\pm1.24^{+1.63}_{-1.18})\times 10^{-6}.
 \en
The Belle result is subsequently confirmed by the BaBar
measurement \cite{BaBar1}:
 \be
 \B(B^+\to f_0(980)K^+\to \pi^+\pi^-K^+)=
 (9.2\pm1.2^{+2.1}_{-2.6})\times 10^{-6}.
 \en
A recent BaBar analysis of $B^\pm\to K^\pm\pi^\mp\pi^\pm$
 gives a very similar result:
$(9.2\pm1.5\pm0.8)\times 10^{-6}$ \cite{BaBarKpipi}. The world
average is then given by \cite{HFAG}
 \be
 \B(B^+\to f_0(980)K^+\to \pi^+\pi^-K^+)=
 (8.49^{+1.35}_{-1.26})\times 10^{-6}.
 \en
BaBar has also measured the neutral mode $B^0\to f_0(980)K^0$ with
the result \cite{BaBar2}
 \be
 \B(B^0\to f_0(980)K^0\to \pi^+\pi^-K^0)=(6.0\pm0.9\pm1.3)\times
 10^{-6}.
 \en
This channel is of special interest as possible indications of New
Physics beyond the Standard Model (SM) may be observed in the
time-dependent $CP$ asymmetries in the penguin-dominated $B$
decays such as $B^0\to f_0(980)K_S$. The mixing-induced
$CP$-violation parameter $S$ is expected to be very close to
$-\sin\beta$ in the SM. The most recent measurements by BaBar and
Belle yield
 \be
 \sin\beta(f_0K_S)=\cases{ 0.95^{+0.23}_{-0.32}\pm0.10 & BaBar \cite{BaBarS} \cr
   -0.47\pm0.41\pm0.08 & Belle \cite{BelleS}. }
 \en
The deviation from $\sin 2\beta=0.726\pm 0.037$ \cite{Ligeti}
measured from $B\to J/\psi K_S$ may hint at a possible New
Physics.

The absolute branching ratios for $B\to f_0K$ depends critically
on the branching fraction of $f_0(980)\to \pi\pi$. For this
purpose, we use the results from the most recent analysis of
\cite{Anisovich-f0}, namely, $\Gamma_{\pi\pi}=64\pm8$ MeV and
$\Gamma_{\rm tot}=80\pm10$ MeV for $f_0(980)$, to obtain
$\B(f_0(980)\to\pi\pi)=0.80\pm0.14$. \footnote{The ratio
$R\equiv\B(f_0\to\pi^+\pi^-)/\B(f_0\to K^+K^-)\approx 7.1$ is
consistent with the result of $R>2.6^{+0.7}_{-0.6}$ inferred from
the Belle measurements of $\B(B^+\to f_0(980)K^+\to
\pi^+\pi^-K^+)$ [see Eq. (\ref{eq:Bellef0K1})] and $\B(B^+\to
f_0(980)K^+\to K^+K^-K^+)<2.9\times 10^{-6}$ \cite{BelleKKK}. }
This leads to
 \be
 \B(B^+\to f_0(980)K^+)\approx(15.9^{+3.8}_{-3.7})\times 10^{-6},
 \non \\
 \B(B^0\to f_0(980)K^0)\approx (11.3\pm3.6)\times 10^{-6}.
 \en
Comparing with the averaged branching ratios, $(12.1\pm 0.8)\times
10^{-6}$ for $B^+\to \pi^0 K^+$ and $(11.5\pm1.0)\times 10^{-6}$
for $B^0\to\pi^0K^0$ \cite{HFAG}, we see that $f_0(980)K^+\gsim
\pi^0 K^+$ and $f_0(980)K^0\approx \pi^0K^0$.

Theoretically, the decay $B\to f_0K$ has been studied in
\cite{Chen1} and \cite{Chen2} within the framework of the pQCD
approach based on the $k_T$ factorization theorem. It is found
that the branching ratio is of order $5\times 10^{-6}$ (see Fig. 2
of \cite{Chen2}), which is smaller than the measured value of
$\B(B^+\to f_0K^+)$ by a factor of 3. In the present paper we
shall re-examine this decay within the QCD factorization approach
\cite{BBNS,BBNS1,BN}.

The underlying structure of the parity-even meson $f_0(980)$ is
still controversial and not clear: It can be a conventional
2-quark $P$-wave state or a $S$-wave meson made of four quarks
$qq\bar q\bar q$. It will be interesting to see if these two
different scenarios for $f_0(980)$ can be tested in $B\to f_0K$
decays, an issue which we will address briefly in this work.

There are two distinct penguin contributions to $B\to f_0K$: one
is related to the $u$ quark component of $f_0$ and the other to
the strange quark content of $f_0$. Owing to the cancellation
between two penguin terms, the former penguin contribution is
severely suppressed, while the latter depends on the unknown
scalar decay constant of $f_0$. Based on the sum rule approach, we
shall show that the magnitude of the $f_0$ scalar decay constant
is sunstantially larger than the corresponding decay constant of
pseudoscalar mesons and hence the $f_0K$ rate can be comparable to
the $\pi^0 K$ one. However, the predicted branching ratio of $B\to
f_0K$ is still lower than experiment by $\sim 45\%$ for the
$f_0K^-$ mode and $\sim 30\%$ for $f_0K^0$. There are several
possible mechanisms such as final-state interactions,
flavor-singlet contributions, large annihilation contributions
that may account for the enhancement of $f_0K$. In the present
work, we will focus on the subleading effects arising from the
three-parton Fock states of the $f_0$ and from the spectator gluon
inside the $B$ meson.

It is commonly assumed that only the valence quarks of the initial
and final state hadrons participate in the decays. Nevertheless, a
real hadron in QCD language should be described by a set of Fock
states for which each state has the same quantum number as the
hadron. For instance,
\begin{eqnarray}\label{eq:fockexpansion}
|B^-\rangle &=& \psi_{b\bar u}^B |b\bar u\rangle + \psi_{b\bar u
g}^B |b\bar u g\rangle + \psi_{b\bar u q\bar q}^B |b\bar u q \bar
q\rangle + \psi_{b\bar u c\bar c}^B |b\bar u c \bar
c\rangle+ \dots\,, \nonumber\\
|f_0\rangle &=&  \psi_{n\bar n}^{f_0} |n\bar n\rangle +
 \psi_{s\bar s}^{f_0} |s\bar s\rangle + \psi_{n\bar n g}^{f_0} |n\bar n g\rangle +
\psi_{s\bar s g}^{f_0} |s\bar s g\rangle + \sum_q (\psi_{n\bar n
q\bar q}^{f_0} |n\bar n q \bar q\rangle + \psi_{s\bar s q\bar
q}^{f_0} |s\bar
s q \bar q\rangle) +\dots\,,\nonumber\\
|K^-\rangle &=& \psi_{s\bar u}^K|s\bar u\rangle + \psi_{s\bar u
g}^K |s\bar u g\rangle + \psi_{s\bar u q\bar q}^K |s\bar u q \bar
q\rangle + \dots\,,
\end{eqnarray}
where $n\bar n\equiv(u\bar u + d \bar d)/\sqrt{2}$. The extra
gluon(s) or quark pair(s) appearing in higher Fock states are the
results of QCD interactions.  The $c \bar c$ pairs due to a single
gluon splitting $g\to c \bar c$, described by the
Gribov-Lipatov-Altarelli-Parisi (GLAP) evolution
equation,\footnote{We should remind the readers that the
light-cone wave functions in Eq.~(\ref{eq:fockexpansion}) are
Lorentz invariant using the light-cone quantization in the
light-cone gauge $A^+=0$~\cite{Brodsky:2001wx}. However, here we
adopt the equal-time quantization and choose the $B$ rest frame.
Each Fock-state wave function of the $B$ meson is no longer boost
invariant. In contrast, the energetic $f_0$ and $K$ produced in
$B$ decays are represented in terms of light-cone distribution
amplitudes (LCDAs) in conformal expansion~\cite{Braun}. Strictly
speaking, the GLAP equation can be applied only to the partons in
the infinite momentum frame or  to the light-cone quantization
framework. }
are basically extrinsic to the bound-state nature of the hadron.
In contrast, the $c\bar c$ pairs, which are entangled through
multiply gluonic interactions to the valence quarks, at least via
$g^* g^* \to c \bar c$, are intrinsic to the hadronic
structure~\cite{Brodsky:2001yt,Chang:2001iy, Gabbiani:2002ti}. It
has been estimated that the intrinsic charm probability in the
proton is $\lesssim 1\%$~\cite{Franz:2000ee}. The intrinsic charm
may  explain the $\rho\pi$ puzzle in $J/\psi (\psi')$
decays~\cite{Brodsky:1997fj}.

The study of the intrinsic charm component in the $B$ meson is
interesting and has attracted a great deal of
attentions~\cite{Brodsky:2001yt,Chang:2001iy, Gabbiani:2002ti}.
Although the extrinsic charm pairs carry only small momentum
fraction of hadrons, it has been argued that the intrinsic $c\bar
c$ pair could share 20\% momentum fraction of the parent $B$
meson~\cite{Chang:2001iy}. As discussed in \cite{Brodsky:2001yt},
the intrinsic charm may give rise to 20\% corrections to $B\to K
\pi$ amplitudes. Consistently, $|\psi_{b\bar u c\bar c}^B/
\psi_{b\bar u}^B|^2$ integrating over the phase space is estimated
to be roughly of order $(20\%)^2=4\%$~\cite{Gabbiani:2002ti},
compared to the charm content $\lesssim 1\%$ in the proton.
Similar to the intrinsic charm case, one may ask how large is the
intrinsic gluon within the $B$ meson. The gluon content of the $B$
meson is intimately related to the energy of light degrees of
freedom (or the so-called ``brown muck"), namely,
$\bar\Lambda=m_B-m_b$. Even though $\bar\Lambda\gtrsim 450$~MeV is
larger than the constituent light quark mass, it may suggest that
the intrinsic gluon within the $B$ meson cannot be neglected. To
estimate the possible contributions originated from the intrinsic
gluon within the $B$ meson, we introduce the three-particle
operator $\bar b \gamma^\alpha g_s \widetilde G_{\alpha\mu} u$,
which can couple to the $B$ meson. Similar definitions for $\pi$
and $\rho$ can be found in the
literature~\cite{Novikov:1983jt,Zhitnitsky:1985us}. Hence, the
intrinsic gluon effects can be estimated. The detailed calculation
is shown in Sec.~\ref{sec:intrinsicgluon}. As a result, we find
that the intrinsic gluon may give rise to 20-40\% corrections to
the decay amplitudes. In contrast, we shall show in
Sec.~\ref{sec:subQCDF} that the extrinsic gluon effect due to the
splitting $ b \to b g$ is negligible, which is described by the
GLAP evolution equation in the infinite momentum frame.

The layout of the present paper is organized as follows. In Sec.
II we shall first study $B\to f_0K$ decays within the framework of
QCD factorization and discuss the internal structure of the
$f_0(980)$. We then continue to explore the subleading effects
arising from the three-parton Fock states of the $f_0$ and from
the spectator gluon inside the $B$ meson. We present numerical
results and discussions in Sec. III and give conclusions in Sec.
IV. Appendices A and B are devoted to the determination of the
scalar decay constant and distribution amplitudes of $f_0$,
respectively, Appendix C outlines the sum-rule calculation for the
decay constant $\delta^2_{f_0^q}$ and Appendix D for the intrinsic
gluon content of the $B$ meson.

\section{$B\to f_0(980)K$ decay amplitudes}

\subsection{$B\to f_0(980)K$ decay amplitudes in QCD factorization of leading Fock states}

\subsubsection{Framework}

 To proceed we first discuss the decay constants and
form factors. The decay constants are defined by
 \be \label{eq:decayc}
 \la K(p)|A_\mu|0\ra=-if_Kp_\mu, \qquad \la f_0|V_\mu|0\ra=0,
 \qquad \la f_0|\bar qq|0\ra=m_{f_0}\bar f_q.
 \en
The scalar meson $f_0(980)$ cannot be produced via the vector
current owing to charge conjugation invariance or conservation of
vector current. The scalar decay constant $\bar f_q$ will be
discussed later. Form factors for $B\to P$ and $B\to S$
transitions ($P$: pseudoscalar meson, $S$: scalar meson) are
defined by \cite{BSW}
 \be \label{m.e.}
 \la P(p_P)|V_\mu|B(p_B)\ra &=& \left(p_{B\mu}+p_{P\mu}-{m_B^2-m_P^2\over q^2}\,q_ \mu\right)
F_1^{BP}(q^2)+{m_B^2-m_P^2\over q^2}q_\mu\,F_0^{BP}(q^2), \non \\
 \en
where $q_\mu=(p_B-p_P)_\mu$, and \cite{CCH}\footnote{As shown in
\cite{CCH}, a factor of $(-i)$ is needed in Eq. (\ref{SPff}) in
order for the $B\to S$ form factors to be positive. This also can
be checked from heavy quark symmetry \cite{CCH}.}
 \be \label{SPff}
 \la S(p_S)|A_\mu|B(p_B)\ra &=& -i\Bigg[\left(p_{B\mu}+
 p_{S\mu}-{m_B^2-m_S^2\over q^2}\,q_ \mu\right)
F_1^{BS}(q^2)  \non \\
&& +{m_B^2-m_S^2\over q^2}q_\mu\,F_0^{BS}(q^2)\Bigg].
 \en

\begin{figure}[t]
\vspace{0cm}
  \centerline{\psfig{figure=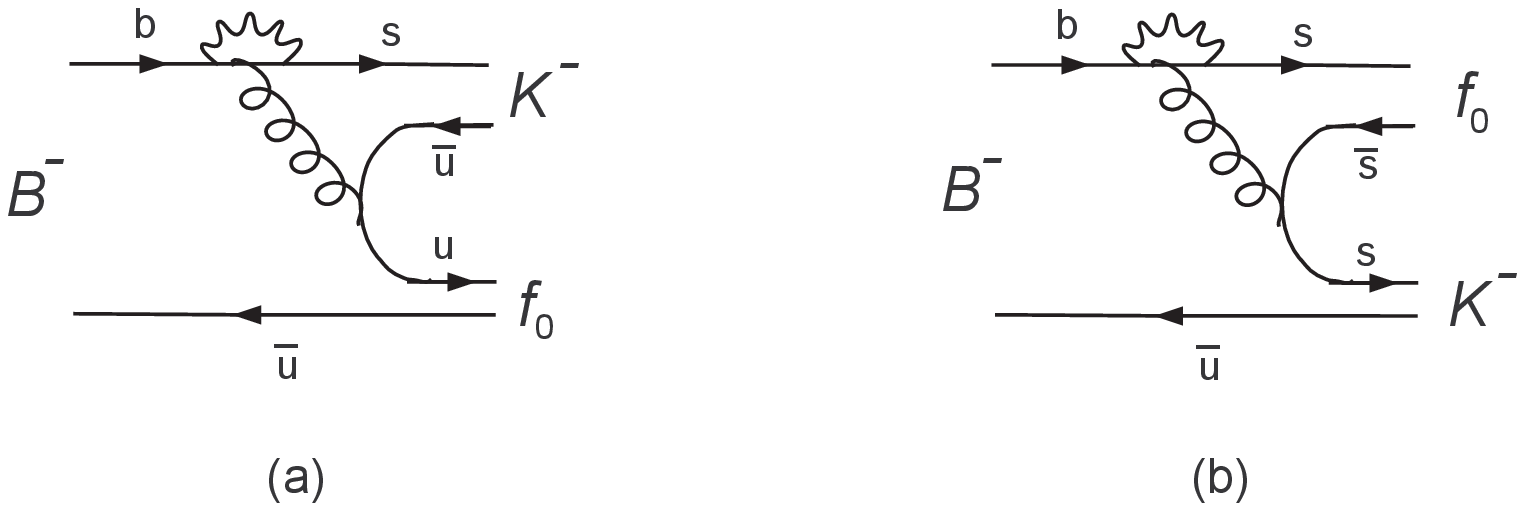,width=10cm}}
\vspace{0cm}
    \caption[]{\small Penguin contributions to $B^-\to
    f_0(980)K^-$in the 2-quark picture for $f_0(980)$.}
    \label{fig:BfK2q}
\end{figure}

The penguin-dominated $B^-\to f_0K^-$ receive two different types
of penguin contributions as depicted in Fig. 1. Within the
framework of QCD factorization \cite{BBNS}, the $B\to f_0K$ decay
amplitudes read\footnote{The relative sign between
$a_6^p(f_0K)f_K$ and $a_6^p(Kf_0)\bar f_s$ terms in the decay
amplitude of $B\to f_0K$ obtained in \cite{Delepine} is opposite
to ours.}
 \be \label{eq:AmpfK}
A(B^- \to f_0 K^- ) &=& -\frac{G_F}{\sqrt{2}}\sum_{p=u,c}\lambda_p
 \Bigg\{ \left[ a_1(f_0K^-)\delta^p_u+(a_4^p-r_\chi^Ka_6^p)(f_0K^-)
 +(a_{10}^p-r_\chi^Ka_8^p)(f_0K^-) \right] \non \\
 &\times& f_KF_0^{Bf_0^u}(m_K^2)(m_B^2-m_{f_0}^2)
 + (2a_6^p-a_8^p)(K^-f_0)\bar f_s\,{m_{f_0}\over
 m_b}F_0^{BK}(m^2_{f_0})(m_B^2-m_K^2)\non \\
 &-& \sum_{q=u,s}f_Bf_K\bar f_q\bigg[b_3^q(f_0
 K^-)+b_{\rm 3EW}^q(f_0K^-)\bigg] \Bigg\}, \non \\
A(\ov B^0 \to f_0\ov K^0 ) &=&
-\frac{G_F}{\sqrt{2}}\sum_{p=u,c}\lambda_p
 \Bigg\{ \left[(a_4^p-r_\chi^Ka_6^p)(f_0\ov K^0)
 -{1\over 2}(a_{10}^p-r_\chi^Ka_8^p)(f_0\ov K^0) \right] \non \\
 &\times& f_KF_0^{Bf_0^u}(m_K^2)(m_B^2-m_{f_0}^2)
 + (2a_6^p-a_8^p)(\ov K^0f_0)\bar f_s\,{m_{f_0}\over
 m_b}F_0^{BK}(m^2_{f_0})(m_B^2-m_K^2)\non \\
  &-& \sum_{q=u,s}f_Bf_K\bar f_q\bigg[b_3^q(f_0\ov
  K^0)-{1\over 2}b_{\rm 3EW}^q(f_0\ov K^0)\bigg] \Bigg\},
 \en
where $\lambda_p\equiv V_{pb}V_{ps}^*$,
$r^K_\chi(\mu)=2m_K^2/[m_b(\mu)(m_u(\mu)+m_s(\mu))]$, and weak
annihilation contributions described by $b_3$ and $b_{\rm 3EW}$
terms will be discussed shortly. In Eq. (\ref{eq:AmpfK}) the
superscript $u$ of the form factor $F_0^{Bf_0^u}$ reminds us that
it is the $u$ quark component of $f_0$ involved in the form factor
transition [see Fig. 1(a)]. In contrast, the subscript $s$ of the
decay constant $\bar f_s$ indicates that it is the strange quark
content of $f_0$ responsible for the penguin contribution of Fig.
1(b).

The effective parameters $a_i^p$ with $p=u,c$ in Eq.
(\ref{eq:AmpfK}) can be calculated in the QCD factorization
approach \cite{BBNS}. They are basically the Wilson coefficients
in conjunction with short-distance nonfactorizable corrections
such as vertex corrections and hard spectator interactions. In
general, they have the expressions \cite{BBNS,BN}
 \be
 a_i^p(M_1M_2) &=& c_i+{c_{i\pm1}\over N_c}
  +{c_{i\pm1}\over N_c}\,{C_F\alpha_s\over
 4\pi}\Big[V_i(M_2)+{4\pi^2\over N_c}H_i(M_1M_2)\Big]+P_i^p(M_2),
 \en
where $i=1,\cdots,10$,  the upper (lower) signs apply when $i$ is
odd (even), $c_i$ are the Wilson coefficients,
$C_F=(N_c^2-1)/(2N_c)$ with $N_c=3$, $M_2$ is the emitted meson
and $M_1$ shares the same spectator quark with the $B$ meson. The
quantities $V_i(M_2)$ account for vertex corrections,
$H_i(M_1M_2)$ for hard spectator interactions with a hard gluon
exchange between the emitted meson and the spectator quark of the
$B$ meson and $P_i(M_2)$ for penguin contractions. The explicit
expressions of these quantities  can be found in \cite{BBNS,BN}.
The hard spectator function $H$ reads
 \be
 H_i(f_0K) &=& {\bar f_u f_B\over
F_0^{Bf_0^u}(0)m^2_B}\int^1_0 {d\rho\over\rho}\,
\Phi_B(\rho)\int^1_0 {d\xi\over \bar\xi} \,\Phi_K(\xi)\int^1_0
{d\eta\over \deta}\left[\Phi_{f_0}(\eta)+{2m_{f_0}\over
m_b}\,{\bar\xi\over \xi}\,\Phi_{f_0}^p(\eta)\right],
 \en
for $i=1,4,10$ and $H_i=0$ for $i=6,8$. where $\bar\xi\equiv
1-\xi$ and $\bar\eta=1-\eta$. As for the parameters
$a_{6,8}^{u,c}(\ov K^0f_0)$ appearing in Eq. (\ref{eq:AmpfK}),
they have the same expressions as $a_{6,8}^{u,c}(f_0\ov K^0)$
except that the penguin function $\hat G_K$ (see Eq. (55) of
\cite{BBNS}) is replaced by $\hat G_{f_0}$ and $\Phi^p_K$ by
$\Phi^p_{f_0}$.

Weak annihilation contributions to $B\to f_0 K$ are described by
the terms $b_3$ and $b_{\rm 3EW}$ in Eq. (\ref{eq:AmpfK}) which
have the expressions
 \be
 b_3^q &=& {C_F\over
 N_c^2}\left[c_3A_1^{i(q)}+c_5(A_3^{i(q)}+A_3^{f(q)})+N_cc_6A_3^{f(q)}\right], \non \\
 b_{\rm 3EW}^q &=& {C_F\over
 N_c^2}\left[c_9A_1^{i(q)}+c_7(A_3^{i(q)}+A_3^{f(q)})+N_cc_8A_3^{i(q)}\right],
 \en
where $A_3^f$ is the factorizable annihilation amplitude induced
from $(S-P)(S+P)$ operator and $A_{1,3}^i$ are nonfactorizable
ones induced from $(V-A)(V-A)$ and $(S-P)(S+P)$ operators,
respectively. It is evident that the dominant annihilation
contribution arises from the factorizable penguin-induced
annihilation characterized by $A_3^f$. Their explicit expressions
are given by (see also \cite{BN})
 \be \label{eq:ann}
 A_1^{i(q)}&=& \pi\alpha_s\int_0^1 dxdy
 \left\{ \Phi_K(x)\Phi^{(q)}_{f_0}(y)\left[{1\over y(1-x\bar y)}+{1\over \bar
 x^2y}\right]-2r_\chi^K{m_{f_0}\over m_b}\Phi_K^p(x)\Phi_{f_0}^{(q)p}(y)\,{2\over \bar
 xy}\right\}, \non  \\
 A_3^{i(q)}&=& \pi\alpha_s\int_0^1 dxdy\left\{ {2m_{f_0}\over
 m_b}\Phi_K(x)\Phi_{f_0}^{(q)p}(x)\,{2\bar y\over \bar xy(1-x\bar
 y)}+r_\chi^K\Phi^{(q)}_{f_0}(y)\Phi^p_K(x)\,{2x\over \bar xy(1-x\bar y)}\right\}, \non \\
 A_3^{f(q)} &=& \pi\alpha_s\int_0^1 dxdy\left\{ {2m_{f_0}\over
 m_b}\Phi_K(x)\Phi_{f_0}^{(q)p}(y)\,{2(1+\bar x)\over \bar x^2y}-r_\chi^K
 \Phi^{(q)}_{f_0}(y)\Phi^p_K(x)\,{2(1+y)\over \bar xy^2}\right\},
 \en
where $q=u,s$, $\bar x=1-x,~\bar y=1-y$, $\Phi_M$ ($\Phi_M^p$) is
the twist-2 (twist-3) light-cone distribution amplitude of the
meson $M$.

Although the parameters $a_i(i\neq 6,8)$ and $a_{6,8}r_\chi$ are
formally renormalization scale and $\gamma_5$ scheme independent,
in practice there exists some residual scale dependence in
$a_i(\mu)$ to finite order. To be specific, we shall evaluate the
vertex corrections to the decay amplitude at the scale
$\mu=m_b/2$. In contrast, as stressed in \cite{BBNS}, the hard
spectator and annihilation contributions should be evaluated at
the hard-collinear scale $\mu_h=\sqrt{\mu\Lambda_h}$ with
$\Lambda_h\approx 500 $ MeV. There is one more serious
complication about these contributions; that is, while QCD
factorization predictions are model independent in the
$m_b\to\infty$ limit, power corrections always involve troublesome
endpoint divergences. For example, the annihilation amplitude has
endpoint divergences even at twist-2 level and the hard spectator
scattering diagram at twist-3 order is power suppressed and posses
soft and collinear divergences arising from the soft spectator
quark. Since the treatment of endpoint divergences is model
dependent, subleading power corrections generally can be studied
only in a phenomenological way. We shall follow \cite{BBNS} to
parametrize the endpoint divergence $X_A\equiv\int^1_0 dx/\bar x$
in the annihilation diagram as
 \be \label{eq:XA}
 X_A=\ln\left({m_B\over \Lambda_h}\right)(1+\rho_A e^{i\phi_A}),
 \en
where $\rho_A$ is a complex parameter $0\leq |\rho_A|\leq 1$.
Likewise, the endpoint divergence $X_H$ in the hard spectator
contributions can be parametrized in a similar way.

Note that $a_4$ and $a_6$ penguin terms contribute constructively
to $\pi^0K^-$ but destructively to $f_0 K^-$. Therefore, the
contribution to $B\to f_0K$ from Fig. 1(a) will be severely
suppressed. The contribution from Fig. 1(b) is suppressed by
$m_{f_0}/m_b$. Hence, it is naively expected that the $f_0K$ rate
is much smaller than the $\pi^0K$ one. However, as we shall see
below, the scale dependent decay constant $\bar f_s$ is much
larger than $f_\pi$. As a consequence, the branching ratio of
$B\to f_0K$ turns out to be comparable to $B\to \pi^0K$.

\subsubsection{Quark structure of $f_0(980)$ and input parameters}

It is known that the underlying structure of scalar mesons is not
well established theoretically (for a review, see e.g.
\cite{Spanier,Godfrey,Close}). It has been suggested that the
light scalars below or near 1 GeV--the isoscalars $f_0(600)$ (or
$\sigma$), $f_0(980)$, the isodoublet $K_0^*(800)$ (or $\kappa$)
and the isovector $a_0(980)$--form an SU(3) flavor nonet, while
scalar mesons above 1 GeV, namely, $f_0(1370)$, $a_0(1450)$,
$K^*_0(1430)$ and $f_0(1500)/f_0(1710)$, form another nonet. A
consistent picture \cite{Close} provided by the data  suggests
that the scalar meson states above 1 GeV can be identified as a
conventional $q\bar q$ nonet with some possible glue content,
whereas the light scalar mesons below or near 1 GeV form
predominately a $qq\bar q\bar q$ nonet \cite{Jaffe,Alford} with a
possible mixing with $0^+$ $q\bar q$ and glueball states. This is
understandable because in the $q\bar q$ quark model, the $0^+$
meson has a unit of orbital angular momentum and hence it should
have a higher mass above 1 GeV. On the contrary, four quarks
$q^2\bar q^2$ can form a $0^+$ meson without introducing a unit of
orbital angular momentum. Moreover, color and spin dependent
interactions favor a flavor nonet configuration with attraction
between the $qq$ and $\bar q\bar q$ pairs. Therefore, the $0^+$
$q^2\bar q^2$ nonet has a mass near or below 1 GeV. This
four-quark scenario explains naturally the mass degeneracy of
$f_0(980)$ and $a_0(980)$, the broader decay widths of
$\sigma(600)$ and $\kappa(800)$ than $f_0(980)$ and $a_0(980)$,
and the large coupling of $f_0(980)$ and $a_0(980)$ to $K\ov K$.

While the above-mentioned four-quark assignment of $f_0(980)$ is
certainly plausible when the light scalar meson is produced in
low-energy reactions, one may wonder if the energetic $f_0(980)$
produced in $B$ decays is dominated by the four-quark
configuration as it requires to pick up two energetic
quark-antiquark pairs to form a fast-moving light four-quark
scalar meson. The Fock states of $f_0(980)$ consists of $q\bar q$,
$q^2\bar q^2$, $q\bar q g$ etc. [see Eq.
(\ref{eq:fockexpansion})]. It is thus expected that the
distribution amplitude of $f_0$ would be smaller in the four-quark
model than in the two-quark picture. Then one will not be able to
explain the observed $B\to f_0(980)K$ decays.

\begin{figure}[t]
\vspace{0cm}
  \centerline{\psfig{figure=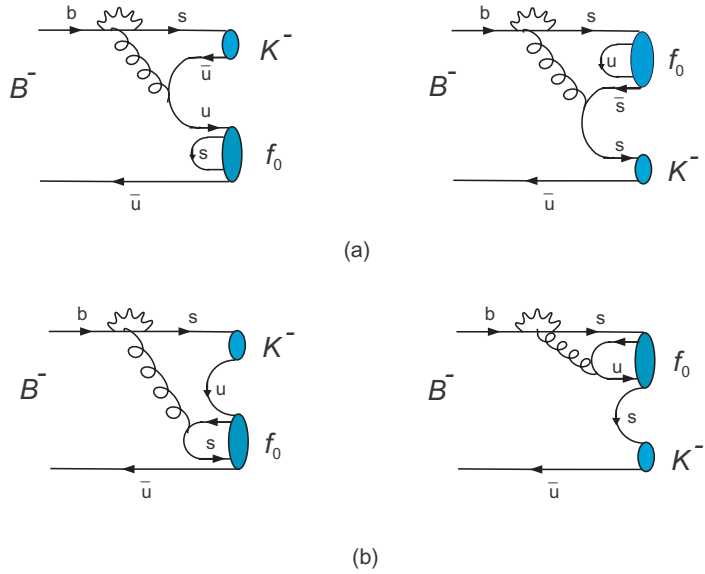,width=10cm}}
\vspace{0cm}
    \caption[]{\small Penguin contributions to $B^-\to
    f_0(980)K^-$in the 4-quark picture for $f_0(980)$.}
    \label{fig:BfK4q}
\end{figure}

Nevertheless, as pointed out in \cite{Brito}, the number of the
quark diagrams for the penguin contributions to $B\to f_0(980)K$
(see Fig. \ref{fig:BfK4q}) in the four-quark scheme for $f_0(980)$
is two times as many as that in the usual 2-quark picture (see
Fig. \ref{fig:BfK2q}). That is, besides the factorizable diagrams
in Fig. \ref{fig:BfK4q}(a), there exist two more nonfactorizable
contributions depicted in Fig. \ref{fig:BfK4q}(b). Therefore, {\it
a priori} there is no reason that the $B\to f_0(980)K$ rate will
be suppressed if $f_0$ is a four-quark state. However, in
practice, it is difficult to give quantitative predictions based
on this scenario as the nonfactorizable diagrams are usually not
amenable. Moreover, even for the factorizable contributions, the
calculation of the $f_0(980)$ decay constant and its form factors
is beyond the conventional quark model, though attempt has been
made in \cite{Brito}. In order to make quantitative calculations
for $B\to f_0(980)K$, we shall assume the conventional 2-quark
description of the light scalar mesons.

In the naive 2-quark picture, $f_0(980)$ is purely an $s\bar s$
state and this is supported by the data of $D_s^+\to f_0\pi^+$ and
$\phi\to f_0\gamma$ implying the copious $f_0(980)$ production via
its $s\bar s$ component. However, there also exist some
experimental evidences indicating that $f_0(980)$ is not purely an
$s\bar s$ state. First, the observation of $\Gamma(J/\psi\to
f_0\omega)\approx {1\over 2}\Gamma(J/\psi\to f_0\phi)$ \cite{PDG}
clearly indicates the existence of the non-strange and strange
quark content in $f_0(980)$. Second, the fact that $f_0(980)$ and
$a_0(980)$ have similar widths and that the $f_0$ width is
dominated by $\pi\pi$ also suggests the composition of $u\bar u$
and $d\bar d$ pairs in $f_0(980)$; that is, $f_0(980)\to\pi\pi$
should not be OZI suppressed relative to $a_0(980)\to\pi\eta$.
Therefore, isoscalars $\sigma(600)$ and $f_0$ must have a mixing
 \be \label{eq:mixing}
 |f_0(980)\ra = |s\bar s\ra\cos\theta+|n\bar n\ra\sin\theta,
 \qquad |\sigma_0(600)\ra = -|s\bar s\ra\sin\theta+|n\bar n\ra\cos\theta,
 \en
with $n\bar n\equiv (\bar uu+\bar dd)/\sqrt{2}$. The distribution
amplitudes $\Phi_s\equiv \Phi_{f_0}^{(s)}$ and $\Phi_n\equiv
\Phi_{f_o}^{(n)}$ corresponding to $f_0^s=\bar ss$ and $f_0^n=\bar
nn$, respectively, are
 \be
 \la f_0^n(p)|\bar q(z)\gamma_\mu q(0)|0\ra &=& p_\mu \tilde f_n\int
 ^1_0 dx e^{ixp\cdot z}\Phi_n(x), \non \\
 \la f_0^s(p)|\bar s(z)\gamma_\mu s(0)|0\ra &=& p_\mu \tilde f_s\int
 ^1_0 dx e^{ixp\cdot z}\Phi_s(x), \non \\
 \la f_0^n(p)|\bar n(z) n(0)|0\ra &=& m_{f_0}^{(n)}\tilde f_n\int
 ^1_0 dx e^{ixp\cdot z}\Phi_n^p(x), \non \\
 \la f_0^s(p)|\bar s(z) s(0)|0\ra &=& m_{f_0}^{(s)}\tilde f_s\int
 ^1_0 dx e^{ixp\cdot z}\Phi_s^p(x),
 \en
with $\tilde f_q$ being defined by
 \be \label{eq:decayc0}
 \la f_0^q|\bar qq|0\ra=m_{f_0}^{(q)}\tilde f_q.
 \en
They satisfy the relations $\Phi_{n,s}(x)=-\Phi_{n,s}(1-x)$ due to
charge conjugation invariance (that is, the distribution amplitude
vanishes at $x=1/2$) and $\Phi_{n,s}^p(x)=\Phi_{n,s}^p(1-x)$ so
that $\int^1_0dx\,\Phi_{n,s}(x)=0$. For the scalar meson made of
$q\bar q$, its general distribution amplitude has the form
 \be
 \Phi_S(x,\mu)=6x(1-x)\left[B_0(\mu)+\sum_{m=1}^\infty
 B_m(\mu)\,C_m^{3/2}(2x-1)\right],
 \en
where $B_m$ are Gegenbauer moments and $C_m^{3/2}$ are the
Gegenbauer polynomials. For the isosinglet scalar mesons $\sigma$
and $f_0$, $B_0=0$ and only the odd Gegenbauer polynomials
contribute. Hence, the light-cone distribution amplitudes (LCDAs)
for $f_0$ read
 \be \label{eq:DAf}
 \Phi_{n,s}(x,\mu)=6x(1-x)\sum_{m=1,3,5,\cdots}B_m^{(n,s)}
 (\mu)C_m^{3/2}(2x-1).
 \en
Using the QCD sum rule technique we have evaluated in Appendix B
the Gegenbauer moments up to $m=5$. In particular, we obtain
$B_1^{(n)}=-0.92\pm0.08$ for $\Phi_n$ and $B_1^{(s)}=0.8B_1^{(n)}$
for $\Phi_s$ at $\mu=1$ GeV. Our results are consistent with the
empirical result of $|B_1|=1.1$ inferred from the analysis in
\cite{Diehl}. As we shall see below, the negative sign of $B_1$ is
indeed strongly favored when the predicted $f_0K$ rates are
confronted with the data. As for the twist-3 distribution
amplitude $\Phi_{f_0}^p(x)$, its asymptotic form is the same as
the light pseudoscalar meson to the leading conformal expansion
\cite{Braun}. Hence, we take
 \be \label{eq:f0DAp}
 \Phi^p_{n,s}(x) =1.
 \en
The asymptotic forms for kaon twist-2 and twist-3 distribution
amplitudes are
 \be \label{eq:DAK}
 \Phi_{K}(x)=6x(1-x), \qquad \Phi_K^p(x)=1.
 \en

In the $q\bar q$ description of $f_0(980)$, it follows from that
 \be
 F_0^{B^-f_0}={1\over\sqrt{2}}\sin\theta\,F_0^{B^-f_0^{u\bar u}}, \qquad
 F_0^{B^0f_0}={1\over\sqrt{2}}\sin\theta\,F_0^{B^0f_0^{d\bar d}},
 \en
where the superscript $q\bar q$ denotes the quark content of $f_0$
involved in the transition. The form factor for $B$ to the scalar
meson transition has been calculated in the covariant light-front
model \cite{CCH}. From Table VI of \cite{CCH}, it is clear that
$F_0^{Bf_0^{q\bar q}}(0)$ with $q\bar q=u\bar u$ or $d\bar d$ is
of order 0.25 which is very similar to $F_0^{B\pi}(0)$. As we
shall see below, a precise estimate of the $B\to f_0(980)$ form
factor is not important because the contribution from Fig.
\ref{fig:BfK2q}(a) is largely suppressed owing to the large
cancellation between the $a_4$ and $a_6$ penguin terms. Based on
the QCD sum-rule method, the scalar decay constant $\tilde f_s$
defined in Eq. (\ref{eq:decayc0}) has been estimated in
\cite{Fazio} and \cite{Bediaga} with similar results, namely,
$\tilde f_s\approx 0.18$ GeV at a typical hadronic scale. Notice
that, in contrast to the pseudoscalar meson decay constants, the
scalar decay constant $\tilde f_q$ is scale dependent. Taking into
account the scale dependence of $\tilde f_q$ and radiative
corrections to the quark loops in the OPE series, in Appendix A we
have made a careful evaluation of the scalar decay constant using
the sum rule approach and found $\tilde f_s(\mu=1\,{\rm
GeV})=0.33\,{\rm GeV}$ and $\tilde f_s(\mu=2.1\,{\rm
GeV})=0.39\,{\rm GeV}$ [see Eqs. (\ref{eq:fs1}) andIn the
two-quark scenario for $f_0(980)$, the decay constants $\bar
f_{n,s}$ are related to $\tilde f_{n,s}$ defined in
Eq.~({\ref{eq:decayc}) via (\ref{eq:fs2.1})] and similar results
for $\tilde f_n$.
 \begin{eqnarray}
 \bar{f}_s=\frac{m_{f_0}^{(s)}}{m_{f_0}}\tilde{f}_s \cos\theta ,
\qquad  \bar{f}_n=\frac{m_{f_0}^{(n)}}{m_{f_0}}\tilde{f}_n
\sin\theta,
\end{eqnarray}
where use has been made of Eqs. (\ref{eq:decayc}),
(\ref{eq:mixing}) and (\ref{eq:decayc0}).

Experimental implications for the $f_0\!-\!\sigma$ mixing angle
have been discussed in detail in \cite{ChengDSP}:
 \be
 J/\psi\to f_0\phi,~f_0\omega~[39] \quad &\Rightarrow &\quad
 \theta=(34\pm6)^\circ~~{\rm or}~~\theta=(146\pm 6)^\circ,  \non \\
 R=4.03\pm0.14~[39] \quad &\Rightarrow &\quad
 \theta=(25.1\pm0.5)^\circ~~{\rm or}~~\theta=(164.3\pm 0.2)^\circ, \non \\
 R=1.63\pm0.46~[39] \quad &\Rightarrow &\quad
 \theta=(42.3^{+8.3}_{-5.5})^\circ~~{\rm or}~~
 \theta=(158\pm 2)^\circ,  \non \\
 \phi\to f_0\gamma,~f_0\to\gamma\gamma~[40] \quad &\Rightarrow &\quad
 \theta=(5\pm5)^\circ~~{\rm or}~~\theta=(138\pm6)^\circ,  \non \\
 {\rm QCD~sum~rules~and}~f_0~{\rm data}~[41] \quad &\Rightarrow &\quad
 \theta=(27\pm13)^\circ~~{\rm or}~~\theta=(153\pm13)^\circ, \non \\
 {\rm QCD~sum~rules~and}~a_0~{\rm data}~[41] \quad &\Rightarrow &\quad
 \theta=(41\pm11)^\circ~~{\rm or}~~\theta=(139\pm11)^\circ,
 \en
where $R\equiv g^2_{f_0K^+K^-}/g^2_{f_0\pi^+\pi^-}$ measures the
ratio of the $f_0(980)$ coupling to $K^+K^-$ and $\pi^+\pi^-$. In
short, $\theta$ lies in the ranges of $25^\circ<\theta<40^\circ$
and $140^\circ<\theta< 165^\circ$. Note that the phenomenological
analysis of the radiative decays $\phi\to f_0(980)\gamma$ and
$f_0(980)\to\gamma\gamma$ favors the second solution, namely,
$\theta=(138\pm 6)^\circ$.\footnote{In the four-quark scenario for
light scalar mesons, one can also define a similar $f_0-\sigma$
mixing angle. It has been shown that $\theta=174.6^\circ$
\cite{Maiani}.}

\subsection{Subleading QCD factorization amplitudes
arising from 3-parton Fock states of the final state
mesons}\label{sec:subQCDF}

 We shall see in Sec. III that the leading QCD
factorization contributions to $B\to f_0K$ are not adequate to
explain the observed large rate of $f_0K^-$ and $f_0\ov K^0$
rates. This leads us to contemplate some possible mechanisms for
enhancement. In this subsection we study the subleading QCD
factorization amplitudes arising from three-parton Fock states of
the $f_0$ or kaon. As will be shown later, these corrections are
of order $1/m_b^2$ and turn out to be negligible. The relevant
diagrams are depicted in Fig.~\ref{fig:3p}. The study of these
corrections is motivated by the following observations: (i) As
shown in~\cite{KCYang}, Fig.~\ref{fig:3p}(b) could give
significant corrections to $B\to \omega K$ and $\omega \pi$ decays
because of the enhancement by the {\it softer} gluon effect
accompanied by the relatively large Wilson coefficients $c_4$ and
$c_6$. (ii) Consider Fig.~\ref{fig:3p}(a) where the emitted meson
is a pseudoscalar~\cite{BBNS} or a vector meson~\cite{KCYang}. Its
total amplitude vanishes due to the mismatch of the $G$-parity
between the emitted meson and the relevant 3-parton operators. In
contrast, the $G$-parities of 3-parton operators do match with the
$f_0$ meson. Moreover, the contributions can be further enhanced
by the large Wilson coefficients $c_4$ and $c_6$.

In the following calculation, we adopt the conventions
$D_\alpha=\partial_\alpha +ig_s T^a A^a_\alpha$,
$\widetilde{G}_{\alpha\beta}=(1/2)
\epsilon_{\alpha\beta\mu\nu}G^{\mu\nu}$ and $\epsilon^{0123}=-1$.
We also use the Fock-Schwinger gauge to express the gluon field
$A_\mu^a$ in terms of $G_{\mu\nu}^a$ for ensuring the
gauge-invariant nature of the results \cite{KCYang,Yang:1993bp}
\begin{eqnarray}
A_{\mu}^a(x)=-\int_0^1 d v\ v G_{\mu\nu}^a(v x)x^\nu.\nonumber
\label{eq:FSgauge}
\end{eqnarray}
 \begin{figure}
\centerline{
        {\epsfxsize3in \epsffile{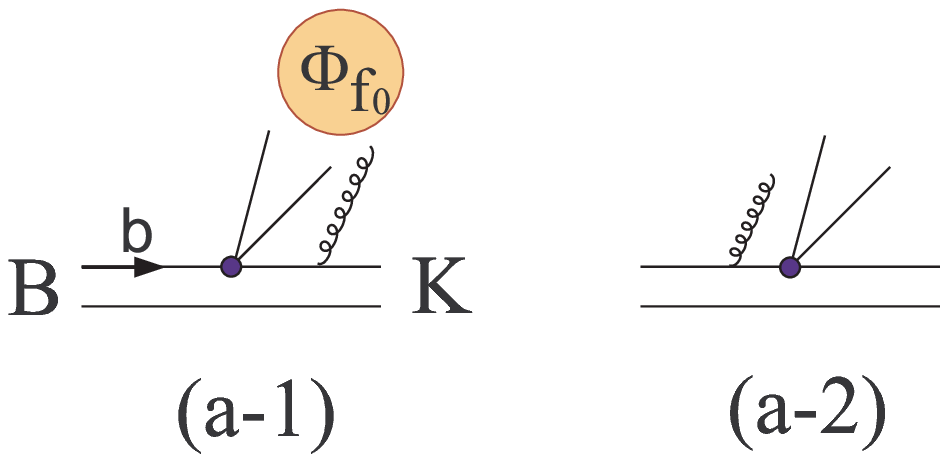}}}
\centerline{
        {\epsfxsize3in \epsffile{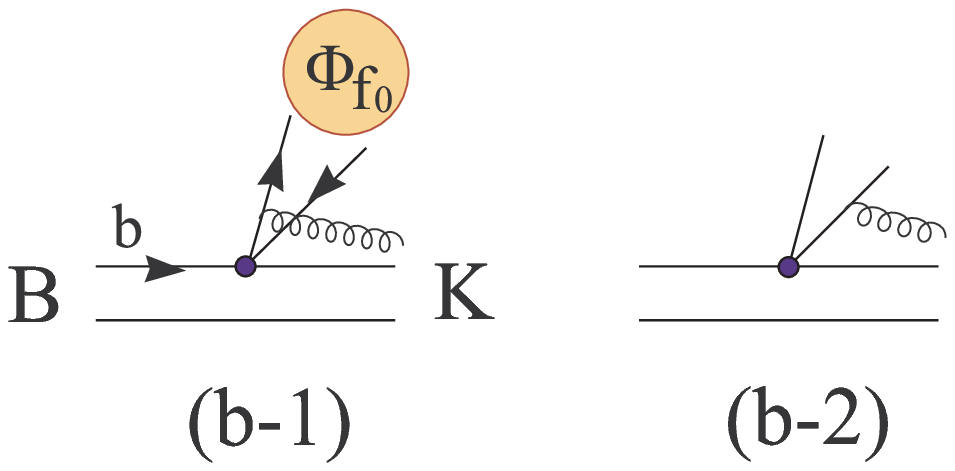}}}
\caption{The subleading QCD factorization amplitudes originating
from  the $\bar qqg$ Fock states of the final state mesons in the
$\overline B \to f_0 \overline K$ decays.}\label{fig:3p}
\end{figure}
The three-parton LCDAs of the $f_0$ meson are defined by
\begin{eqnarray}
\lefteqn{ \langle f_0(p) |\bar{q}(x) g_s
G_{\mu\nu}(vx)\sigma_{\alpha\beta} q(0)|0\rangle }
\nonumber\\
&=& - f_{3 f_0}^{q}[p_\beta (p_\mu g_{\nu\alpha}-p_\nu
g_{\mu\alpha}) - p_\alpha(p_\mu g_{\nu\beta}-p_\nu g_{\mu\beta})]
 \int{\cal D}\alpha\,\phi_{3 f_0}^q e^{ipx(\alpha_u+v\alpha_g)}\,,
\label{3pidefinition}
 \end{eqnarray}
\begin{eqnarray} \lefteqn{ \langle f_0(p) |\bar{q}(x)\gamma_\mu
g_sG_{\alpha\beta}(vx)q(0)|0\rangle} \nonumber
\\
&=& - i \bar{f}_q\left[ p_\beta\left( g_{\alpha\mu}-\frac{x_\alpha
p_\mu}{px}\right) - p_\alpha\left(g_{\beta\mu}-\frac{x_\beta
p_\mu}{px}\right)\right] \int{\cal D}\alpha \phi_\perp^q
(\alpha_i)e^{ipx(\alpha_u+v\alpha_g)}\hspace{1.5cm}{} \nonumber
\\
&& {}- i\bar{f}_q\frac{p_\mu}{px}(p_\alpha x_\beta -p_\beta
x_\alpha ) \int{\cal D}\alpha \,\phi_\parallel^q
(\alpha_i)e^{ipx(\alpha_u+v\alpha_g)}\,, \label{30} \end{eqnarray}
 \begin{eqnarray}
&& \langle f_0(p) |\bar{q}(x)\gamma_\mu \gamma_5
g_s\widetilde{G}_{\alpha\beta}(vx)q(0)|0\rangle= \bar{f}_q \left(
p_\alpha g_{\beta\mu} - p_\beta
g_{\alpha\mu} \right) \nonumber\\
&& \times \int{\cal D}\alpha\,\widetilde{\phi}_\perp^q
e^{ipx(\alpha_u+v\alpha_g)} - \bar{f}_q\frac{p_\mu}{px}(p_\alpha
x_\beta -p_\beta x_\alpha) \int{\cal D}\alpha\,
(\widetilde{\phi}_\parallel^q + \widetilde{\phi}_\perp^q )
e^{ipx(\alpha_u+v\alpha_g)}\,, \label{qqgpi}
\end{eqnarray}
where ${\cal D}\alpha= d\alpha_{\bar q} d\alpha_q d\alpha_g
\delta(1-\alpha_{\bar q}-\alpha_q -\alpha_g)$ with $\alpha_{\bar
q},\alpha_q,\alpha_g$ being the fractions of the $f_0$ momentum
carried by the $\bar q$-quark, $q$-quark and gluon, respectively.
Here $\phi_{3 f_0}^q$ is a twist-3 LCDA, and $\phi_\perp^q$,
$\phi_\parallel^q$, $\widetilde{\phi}_\perp^q$
$\widetilde{\phi}_\parallel^q$ are twist-4 ones given by
\begin{eqnarray}
\phi_{3 f_0}^q (\alpha_i)&=&360 \alpha_q\alpha_{\bar
q}\alpha_g^{2} \Bigg[1+\omega_{1,0}\frac12(7\alpha_g-3)
{}+\omega_{2,0}(2-4\alpha_q\alpha_{\bar q} -
8\alpha_g+8\alpha_g^{2}) \nonumber\\
&&{}+\omega_{1,1}(3\alpha_q \alpha_{\bar q} - 2\alpha_g +
3\alpha_g^{2})\Bigg]~,\nonumber\\
\phi_\perp^q (\alpha_i)&=& 30 \delta_{f^q_0}^2
\alpha_g^{2}(1-\alpha_g)\Bigg[\frac{1}{3}+2 \varepsilon
(1-2\alpha_g)\Bigg] ~, \nonumber
\\
\phi_\parallel^q (\alpha_i)&=&-120 \delta_{f^q_0}^2
\alpha_q\alpha_{\bar q}\alpha_g \Bigg[\frac{1}{3}+ \varepsilon
(1-3\alpha_g) \Bigg],\nonumber\\
 \widetilde{\phi}_\perp^q (\alpha_i)&=&30 \delta_{f^q_0}^2
(\alpha_q-\alpha_{\bar q})\alpha_g^2\Bigg[\frac{1}{3}+2
\varepsilon (1-2\alpha_g)\Bigg] ~, \nonumber
\\
\widetilde{\phi}_\parallel^q (\alpha_i)&=&120 \delta_{f^q_0}^2
\varepsilon (\alpha_q-\alpha_{\bar q})\alpha_q\alpha_{\bar q}
\alpha_g~,\label{eq:3pdas}
\end{eqnarray}
in the conformal expansion (for a further investigation, see
\cite{Braun}). We obtain $\varepsilon\simeq 0.30$, similar to the
case of the pion, and
 \begin{eqnarray} && \delta_{f_0^s}^2(\mu=2.1~{\rm
GeV}) = (0.08\pm 0.01)~{\rm GeV}^2\,,
\nonumber\\
&& \delta_{f_0^n}^2(\mu=2.1~{\rm GeV}) = (0.09\pm 0.01)~{\rm
GeV}^2\,,
\end{eqnarray}
for $\bar qq= \bar ss$ or $\bar qq=\bar nn$. A detailed
calculation of $\delta_{f_0^q}^2$ is shown in
Appendix~\ref{app:c}. Consider the four-quark operator $\bar
s_\alpha\gamma_\mu(1-\gamma_5) b_\beta\ \bar q_\beta \gamma^\mu
(1\mp\gamma_5)q_\alpha$ in Fig.~\ref{fig:3p}, where $\alpha$ and
$\beta$ are color indices. The result of Fig.~\ref{fig:3p}(a.1) is
found to be
\begin{eqnarray}\label{eq:f03fock1}
{\rm Fig.~\ref{fig:3p}(a.1)} =&& -\frac{1}{3}\int^1_0 dv \ v \int
d^4x \int \frac{d^4k}{(2\pi^4)} \langle f_0|\bar q(0) \gamma^\mu
g_s (1\mp \gamma_5)
G_{\alpha\nu}(v x) q(0)|0\rangle x^\nu \nonumber\\
 \times&& \langle \overline K|\bar s(0)
\gamma^\alpha \frac{\not\!k}{k^2}\gamma_\mu (1-\gamma_5)
e^{i(p_K-k) x}  b(0)|\overline B\rangle \\
= &&\frac{2 \bar{f}_q}{3} \langle \overline K|\bar s \not\!p_{f_0}
(1-\gamma_5) b|\overline B\rangle \int D\alpha \frac{1}{\alpha_g
m_B^2}[(\phi_\parallel^q + 2\phi_\perp^q ) \pm
2(\widetilde\phi_\parallel^q - 2\widetilde\phi_\perp^q )],
\nonumber
\end{eqnarray}
where all the components of the coordinate $x$ should be taken
into account in the calculation before the collinear approximation
is applied. Hence, in Eq.~(\ref{eq:f03fock1}), the exponent
arising from $G_{\alpha\nu}(v x)$ is actually $e^{ik^g \cdot xv}$,
where $k^g$ is the $gluon$'s momentum, and the resultant
calculation can be easily performed in the momentum space with the
substitution of $x^\nu \to -(i/v)(\partial/\partial k^g_{\mu})$.
Following the same lines, the result for Fig.~\ref{fig:3p}(a.2)
reads
\begin{eqnarray}
{\rm Fig.~\ref{fig:3p}(a.2)} =&& -\frac{1}{3}\int^1_0 d\alpha\
\alpha\int d^4x \int \frac{d^4k}{(2\pi^4)} \langle f_0|\bar q(0)
\gamma^\mu g_s (1\mp \gamma_5)
G_{\alpha\nu}(\alpha x) q(0)|0\rangle x^\nu \nonumber\\
 \times&& \langle K^-|\bar s(0)
 \gamma_\mu (1-\gamma_5)
\frac{i(\not\!k+m_b)}{k^2-m_b^2}\gamma^\alpha e^{-i(p_B-k) x} b|B^-\rangle \\
= &&\frac{2 \bar{f}_q}{3} \langle K^-|\bar s \not\!p_{f_0}
(1-\gamma_5) b|B^-\rangle \int D\alpha \frac{1}{\alpha_g
m_B^2}[(\phi_\parallel^q + 2\phi_\perp^q ) \mp
2(\widetilde\phi_\parallel^q - 2\widetilde\phi_\perp^q
)],\nonumber
\end{eqnarray}
where we have used the following trick
\begin{eqnarray}
\frac{\not\! p_{f_0} x^2}{ p_{f_0} x} \cong x_+ \gamma_- \to
2i\gamma_-  \frac{\partial}{\partial p_{B-}}
\end{eqnarray}
during the course of calculation and introduced two light-like
vectors: $n_-^\mu$ ($n_-^2=0$), parallel to the momentum of $f_0$,
and $n_+^\mu$ ($n_+^2=0, n_+\cdot n_-=2$), so that $p_{f_0}^\mu
\simeq (p_{f_0}\cdot n_+)n_-^\mu =p_{f_0+}n_-^\mu, x^2\simeq x_-
x_+, p_{f_0}\cdot\gamma=p_{f_{0}+} \gamma_-$.

In Fig.~\ref{fig:3p}(b), the emitted gluon becomes a parton of the
kaon. To proceed, we first take $G_{\mu\nu}(v x)\simeq
G_{\mu\nu}(0)e^{iv k_g^K \cdot x}$ and then adopt the collinear
approximation $k_g^K=\langle\alpha_g\rangle p_K$ in the final
stage, where $\langle\alpha_g\rangle$ is the averaged fraction of
the kaon's momentum carried by the gluon. The calculation is
straightforward and leads to
\begin{eqnarray}
 {\rm Fig.~\ref{fig:3p}(b)}
 &=& \frac{1}{4 N_c} \int_0^1 dv \int_0^1 du
 \Big(\sqrt{2}\bar{f}_n \Phi_{n}(u)+\bar{f}_s\Phi_s(u) \Big)
 \langle K^-|\bar d \gamma_\mu (1-\gamma_5) g_s G_{\nu\beta} b|B^-\rangle
\nonumber\\ && \times i\frac{\partial}{\partial{k_{g\,\beta}^K}}
\bigg\{ {\rm Tr} \bigg[ \not\! p_{f_0} \bigg(
 \frac{\gamma^\nu (v \not\! k_g^K + u \not\! p_{f_0}) \gamma^\mu}
 {(v k_g^K +up_{f_0})^2}
 - \frac{\gamma^\mu (v \not\! k_g^K + \bar u \not\! p_{f_0}) \gamma^\nu}
 {(v k_g^K + \bar up_{f_0})^2} \bigg) \bigg] \bigg\}\nonumber\\
&& \cong 0,\label{eq:3pfig2}
\end{eqnarray}
where we have used the fact that $ \Phi_{n}(\bar u)=-\Phi_n (u),
\Phi_s(\bar u)=-\Phi_s(u)$ with $\bar u=1-u$.

In short, the perturbative contributions coming from higher Fock
states of the final state mesons to the decay amplitudes are (in
units of $G_F/\sqrt{2}$)
\begin{eqnarray} \label{qcdfamp}
 A(B^- \to  f_0 K^-)_{{\rm Fig.}\ref{fig:3p}} &=&
 A(\overline B^0 \to f_0  \overline K^0)_{{\rm Fig.}\ref{fig:3p}}
 \nonumber\\
& =& \frac{4}{3}(m_B^2-m_{K}^2)F_{0}^{BK}(m_{f_0}^2) \int {\cal
D}\alpha \frac{1}{\alpha_g m_B^2} \Bigg\{ V_{ub} V^{*}_{us} c_1
\frac{\bar{f}_n}{\sqrt{2}}
(2\phi_\perp^{n}+\phi_\parallel^{n})\nonumber \\
&& -V_{tb} V^{*}_{ts} \Bigg[(c_4 + c_6)
\Big(\sqrt{2}\bar{f}_n(2\phi_\perp^{n}+\phi_\parallel^{n})+
\bar{f}_s (2\phi_\perp^{s}+\phi_\parallel^{s}) \Big)\nonumber\\
&& +\frac{1}{2}(c_8 + c_{10})
\Big(\frac{\bar{f}_n}{\sqrt{2}}(2\phi_\perp^{n} +
\phi_\parallel^{n}) - \bar{f}_s
(2\phi_\perp^{s}+\phi_\parallel^{s}) \Big) \Bigg]\Bigg\}.
\end{eqnarray}
However, the above result is numerically negligible. Note that
here we do not consider the contributions from the operators
$O_{5,7,8}$; they are not only color suppressed but also of order
$1/m_b^3$.

\subsection{Soft non-factorizable contributions arising from
the intrinsic gluon inside the $B$
meson}\label{sec:intrinsicgluon}

The marginal result of Fig.~\ref{fig:3p}(a.2) with the $b$
splitting, namely,  $b\to bg$ which is governed by the GLAP
evolution equation if $B$ is boosted to the infinite momentum
frame, implies that the extrinsic gluon effect is negligible. In
this subsection, we shall study the intrinsic gluon contributions
to the decay amplitudes. The leading diagram is displayed in
Fig.~\ref{fig:igsr1}(a). Here the intrinsic gluon plays the
spectator role so that this contribution cannot be perturbatively
calculated. Before proceeding, four remarks are in order. (i) The
effects for intrinsic gluon being collinear with the spectator
quark have already been considered in the transition form factor.
(ii) The possible effect that the gluon content of $f_0$ is
produced from the spectator quark is included in the present study
since the spectator quark is a soft object and therefore the gluon
could be one of the constituents of the $B$ meson. (iii) As
illustrated in the previous section,  the extrinsic gluon effect
is quite small and hence there is no double counting problem. (iv)
The amplitude of finding an intrinsic gluon within the $B$ meson
due to the quantum fluctuations are suppressed by $\bar\Lambda
/m_b$. It should be stressed that the idea of twist expansion for
wave functions is not suitable in the present case [see
Fig.~\ref{fig:igsr1}(a)]. Since there is no hard part ready for a
perturbative calculation, the argument for the suppression by
higher twist distribution amplitudes is no longer justified.
Although Fig.~\ref{fig:igsr1}(a) cannot be calculated directly, we
can turn to Fig.~\ref{fig:igsr1}(b) where the $B$ meson state is
replaced by a current operator with large $-p^2$ flowing through.
In this way, Fig.~\ref{fig:igsr1}(b) can be perturbatively
calculated and is related to Fig.~\ref{fig:igsr1}(a) via the
reduction formula and quark-hadron duality.

\begin{figure}
\centerline{
        {\epsfxsize4in \epsffile{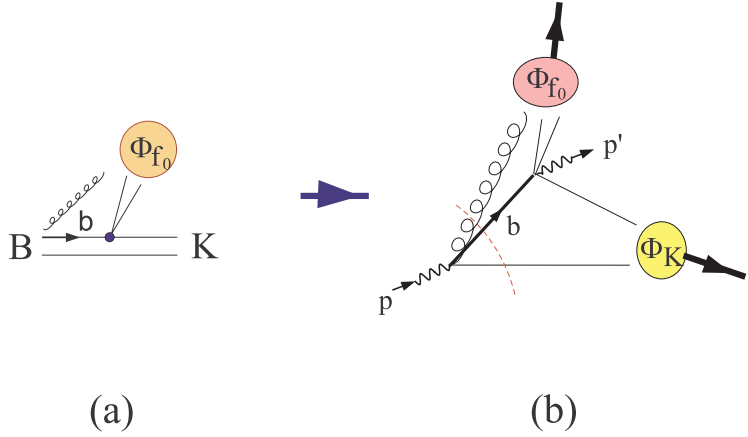}}}
\caption{(a) The contribution to the $B \to f_0 K$ decays arising
from the intrinsic gluon within the $B$ meson  and (b) the
diagrammatic illustration to the correlation function given in
Eq.~(\ref{eq:hadron}).}\label{fig:igsr1}
\end{figure}

Let us embark on the calculation. Consider the four-quark operator
$\bar s_\alpha\gamma_\mu(1-\gamma_5) b_\beta\ \bar q_\beta
\gamma^\mu (1\mp\gamma_5)q_\alpha$ in Fig.~\ref{fig:igsr1}, where
$\alpha$ and $\beta$ are color indices. The result of
Fig.~\ref{fig:igsr1}(a)
 can be represented as
\begin{eqnarray}
&& \langle f_0 K^-|\bar s_\alpha\gamma_\mu(1-\gamma_5) b_\beta\
\bar q_\beta \gamma^\mu (1\mp\gamma_5)q_\alpha|B_g^-\rangle_{\rm
Fig.~\ref{fig:igsr1}(a)} =\langle f_0
K^-|2O^{8(q)}_\mp|B_g^-\rangle_{\rm
Fig.~\ref{fig:igsr1}(a)}\,,\label{eq:figx}
\end{eqnarray}
where
\begin{eqnarray}
O^{8(q)}_\mp= \bar s \gamma_\mu(1-\gamma_5) T^a b\ \bar q
\gamma^\mu(1\mp \gamma_5) T^a q,
\end{eqnarray}
and we have added a subscript ``$g$" to $B^-$ to emphasize the
intrinsic gluon for this case. This matrix element can be
calculated by considering the following correlation function:
 \begin{eqnarray}
\Pi^{\rm IG}_\mu (p,q)&=&i\int d^4x e^{-ipx} \langle K(p_K),
f_0(p_{f_0})| T(2O_\mp^{8}(0),
i\bar b(x)  g_s \widetilde G_{\alpha\mu}(x)\gamma^\alpha u(x)|0\rangle\nonumber\\
&=& \Pi^{\rm IG}(p^2,p'^2) p_\mu + \cdots \,, \label{eq:hadron}
\end{eqnarray}
where the ellipses (and the following one) denote terms with the
structures $(p+p_K)^\mu$ and $(p+p_{f_0})^\mu$, respectively,
which are not relevant to the present study, and the superscript
``IG" stands for intrinsic gluon.

In the deep Euclidean region of $p^2$, as depicted in
Fig.~\ref{fig:igsr1}(b), the correlation function can be
perturbatively calculated in QCD and expressed in terms of LCDAs
of the kaon and $f_0$,
\begin{eqnarray}\label{eq:piqcd}
\Pi_\mu^{\rm IG (QCD)} &=& -{1\over 6}
\epsilon_{\alpha\beta\rho\mu} \int d^4x e^{-ipx} \langle
f_0(p_{f_0})| \bar q(0)\gamma^\nu (1\mp \gamma_5)
G^{\alpha\beta}(x) q(0)|0\rangle \nonumber\\
&&\times \langle K(p_K)|\bar s(0) \gamma_\nu (1-\gamma_5)\int
\frac{d^4k}{(2\pi)^4} e^{ikx} \frac{i(\not\!
k-m_b)}{k^2-m_b^2}\gamma^\rho u(x)|0\rangle \nonumber\\
&=& \frac{1}{3}\bar{f}_{q} f_K \int du \phi_K(u) \int D\alpha
\frac{m_b^2}{m_b^2-(p-up_K -\alpha_g p_{f_0})^2}\nonumber\\
&&\times \Bigg[{1\over 2}(\phi_\perp^{q}+
\phi_\parallel^{q})-(1-\alpha_g)\phi_\perp^{q}\Bigg]p_\mu +
\cdots. \label{eq:quark}
\end{eqnarray}
To estimate  $\langle f_0K|2 O^{8(q)}_\mp|B_g\rangle$, we apply
the quark-hadron duality to the correlation function. Then the
ground state contribution to the correlation function can be
written in the dispersive representation as
\begin{equation}\label{eq:sr2sides}
\int_{m_b^2}^{s_0} ds\,\frac{{\rm Im} \Pi^{\rm IG
(phys)}(s,0)}{s-p^2} = \int_{m_b^2}^{s_0} ds\,\frac{{\rm Im}
\Pi^{\rm IG (QCD)}(s,0)}{s-p^2},
\end{equation}
where $s_0$ is the threshold of higher resonances. On the left
hand side of above equation, the spectral density ${\rm Im}
\Pi^{\rm IG (phys)}(s,0)$ is given by hadronic contributions and
reads
\begin{eqnarray}\label{eq:srphys}
{\rm Im} \Pi^{\rm IG (phys)}(s,0) = f_B\delta^2_B  \langle
f_0K|2O^{8}_\mp|B_g\rangle \delta(s-m_B^2) + {\rm higher\
resonance\ states},
\end{eqnarray}
where $\langle B^-(p_B) |i\bar b \gamma^\alpha g_s \widetilde
G_{\alpha\mu} u|0 \rangle = f_B\delta^2_B p_{B\mu}$. On the right
hand side of Eq.~(\ref{eq:sr2sides}), the spectral density ${\rm
Im} \Pi^{\rm IG (QCD)}(s,0)$ can be easily obtained from
Eq.~(\ref{eq:piqcd}).

Equating Eqs.~(\ref{eq:hadron}) and (\ref{eq:quark}) and
performing the Borel transformation yield
\begin{eqnarray} &&{\bf B} \Bigg[ \frac{1}{m_{B}^{2} - p^{2}} \Bigg]= \exp{\Bigg(-
\frac{m_{B}^{2}}{M^{2}}\Bigg) }, \nonumber \\
&&{\bf B}\Bigg[\frac{1}{ m_{b}^{2} - (p-up_K-\alpha_g
p_{f_0})^{2}} \Bigg]= \frac{1}{\bar\alpha_g} \exp{\Bigg( -
\frac{1+ u\bar\alpha_g m_b^{2}}{ \bar\alpha_g M^{2}}\Bigg)} \,,
 \end{eqnarray}
It leads to the light-cone sum rule
\begin{eqnarray}
&&\langle f_0K|2 O^{8(q)}_\mp|B_g\rangle\nonumber\\
&&=\frac{f_K \bar{f}_{q} m_b^2}{3 f_B \delta_B^2}  \int^1_0 du
\phi_K(u) \int^1_0 d\bar\alpha_g \Theta \Big(
\bar\alpha_g-\frac{m_b^2}{s_0-um_b^2} \Big)\int^{\bar\alpha_g}_0
d\alpha_q\ {1\over \bar\alpha_g}
e^{(m_B^2-\frac{1+u\bar\alpha_g}{\bar\alpha_g}m_b^2)/M^2} \nonumber\\
&&\times \Bigg[\frac{1}{2}(\phi_\perp^q
+\phi_\parallel^q)-\bar\alpha_g \phi_\perp^q \Bigg].
\end{eqnarray}
In Fig.~\ref{fig:igsr2}, we plot $\langle f_0K|2
O^{8(q)}|B\rangle$ as a function of the Borel mass squared. The
relevant normalization scale for the present process is of order
$\sqrt{m_B^2-m_b^2}\approx 2.1$~GeV. Using the parameters
$\widetilde{f}_q,\delta_{f^q_0}^2, \delta_B^2$ given in
Eqs.~(\ref{eq:fs2.1}), (\ref{app:parameter3f2.1}),
(\ref{app:parameterintrinsic}), $f_K=160~{\rm MeV}, f_B=185~{\rm
MeV}$, $m_b=4.85$~GeV for the pole mass of the $b$ quark, and
$s_0=34\pm 1$~GeV$^2$~\cite{Yang:1997nh}, we obtain
\begin{eqnarray}\label{eq:o8}
&&\langle f_0K|2 O^{8(s)}_\mp|B_g\rangle = (-0.20\pm 0.06)\cos\theta, \nonumber\\
&&\langle f_0K|2 O^{8(n)}_\mp|B_g\rangle = (-0.23\pm
0.06)\sin\theta,
\end{eqnarray}
in the Borel window 9~GeV$^2 < M^2 <
12$~GeV$^2$~\cite{Yang:1997nh}. Note that since $\langle f_0K|2
O^{8(q)}_-|B_g\rangle = \langle f_0K|2 O^{8(q)}_+|B_g\rangle $, we
abbreviate $O^{8(q)}_\mp$ as $O^{8(q)}$ in the ensuring study. In
summary, the total contributions arising from the intrinsic gluon
in the $B$ meson to the decay amplitudes are (in units of
$G_F/\sqrt{2}$)
\begin{eqnarray} \label{eq:totIG}
 A(B^- \to  f_0 K^-)_{{\rm Fig.}\ref{fig:igsr1}} &=&
 A(\overline B^0 \to f_0  \overline K^0)_{{\rm Fig.}\ref{fig:igsr1}}
 \nonumber\\
& =& -V_{tb} V^{*}_{ts} \Bigg[(c_4 + c_6) \Big(\sqrt{2}\langle
f_0K|2 O^{8(n)}|B_g\rangle
+ \langle f_0K|2 O^{8(s)}|B_g\rangle \Big)\nonumber\\
&& +\frac{1}{2}(c_8 + c_{10}) \Big(\frac{1}{\sqrt{2}}\langle
f_0K|2 O^{8(n)}|B_g\rangle - \langle f_0K|2 O^{8(s)}|B_g\rangle
\Big) \Bigg].
\end{eqnarray}
\begin{figure}[t]
\vspace{0cm} \centerline{\epsfig{file=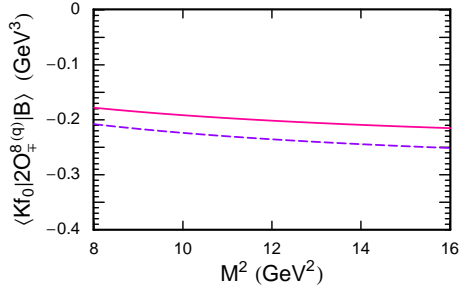,clip=46mm,
width=6cm} } \vspace{0cm} \caption{\small $\langle f_0K|2
O^{8(q)}_\mp|B_g\rangle$ as a function of the Borel mass squared.
The solid curve stands for $\langle f_0K|2
O^{8(s)}_\mp|B_g\rangle$, while the dashed curve for $\langle
f_0K|2 O^{8(n)}_\mp|B_g\rangle$.} \label{fig:igsr2}
\end{figure}

\section{Results and Discussions}
To proceed the numerical calculations, we shall follow
\cite{BBNS,BN} for the choices of the relevant parameters needed
for QCDF calculations except for the form factors and CKM matrix
elements. For form factors we shall use those derived in the
covariant light-front quark model \cite{CCH}. For CKM matrix
elements we use the Wolfenstein parameters $A=0.801$,
$\lambda=0.2265$, $\bar\rho=0.189$ and $\bar\eta=0.358$
\cite{CKMfitter}. For endpoint divergences encountered in hard
spectator and annihilation contributions we take the default
values $\rho_A=\rho_H=0$ [see Eq. (\ref{eq:XA})]. For the running
current quark masses we employ
 \be
 && m_b(m_b)=4.2\,{\rm GeV}, \qquad m_b(2.1\,{\rm GeV})=4.95\,{\rm
 GeV}, \qquad m_b(1\,{\rm GeV})=6.89\,{\rm
 GeV}, \non \\
 && m_c(m_b)=1.3\,{\rm GeV}, \qquad m_c(2.1\,{\rm GeV})=1.51\,{\rm
 GeV}, \non \\
 && m_s(2.1\,{\rm GeV})=90\,{\rm MeV}, \qquad m_s(1\,{\rm GeV})=119\,{\rm
 MeV},
 \en
and $m_q(\mu)/m_s(\mu)=0.0413$ \cite{BN}. The strong coupling
constants are given by \cite{alphas}
 \be
 \alpha_s(2.1\,{\rm GeV})=0.293,\qquad\quad \alpha_s(1\,{\rm
 GeV})=0.489\,.
 \en

It is ready to perform numerical calculations. At the scale
$\mu=2.1$ GeV, the numerical results for the relevant
$a_i^p(f_0K)$ and $a_{6,8}^p(Kf_0)$ are
 \be \label{eq:ai}
  &&  a_4^u = -0.0533 - i0.0183, \qquad\qquad
  a_4^c = -0.0610 - i0.0064, \non \\
  && a_6^u = -0.0568 -i 0.0163, \qquad\qquad
  a_6^c = -0.0612 -i 0.0039, \non \\
  && a_8^u = (74.4 - i4.5)\times 10^{-5}, \qquad\quad~
  a_8^c = (73.6 - i2.3)\times 10^{-5},  \non \\
  && a_{10}^u = (561+i 90  )\times 10^{-5}, \qquad\quad~~
  a_{10}^c = (560 + i92 )\times 10^{-5}, \non \\
  && a_1 = 1.337 + i0.0288, \qquad\qquad\quad~~ a_{6,8}^p(Kf_0) =
  a_{6,8}^p(f_0K).
  \en
Note that the effective Wilson coefficients $a_1,a_4$ and $a_{10}$
receive large contributions from the hard spectator interactions
$H(f_0K)$. Consequently, $a_4$ and $a_6$ are similar. Using the
distribution amplitudes of the kaon and $f_0(980)$ given in Eqs.
(\ref{eq:DAK}) and (\ref{eq:DAf}), respectively, the annihilation
contributions shown in Eq. (\ref{eq:ann}) can be simplified to
 \be
 {\cal A}_1^{i(q)} &\approx& \pi\alpha_s\left[ -18 B_1^{(q)}(X_A+32-{3\over
 2}\pi^2)-4r_\chi^K{m_{f_0}
 \over m_b}X_A^2\right], \non \\
 {\cal A}_3^{i(q)} &\approx& 6\pi\alpha_s\left[ 2{m_{f_0}\over
 m_b}(X_A^2-2X_A+{1\over 3}\pi^2)-3r_\chi^K B_1^{(q)}(X_A^2-4X_A+4)\right], \non \\
 {\cal A}_3^{f(q)} &\approx& 6\pi\alpha_s X_A\left[ 2{m_{f_0}\over
 m_b}(2X_A-1)+r_\chi^K B_1^{(q)}(6X_A-11)\right],
 \en
where the endpoint divergence $X_A$ is defined in Eq.
(\ref{eq:XA}) and only the first Gegenbauer polynomial in the
distribution amplitudes is kept for simplicity.

Shown in Fig. \ref{fig:BRqcdf} are the branching ratios of $B^-\to
f_0(980)K^-$ and $B^0\to f_0(980)K^0$ versus the
strange-nonstrange mixing angle $\theta$ of $f_0(980)$. It is
evident that the coefficient $B_1^{(q)}$ appearing in the
distribution amplitude of $f_0$ [see Eq. (\ref{eq:DAf})] is
preferred to be negative so that the annihilation terms make
constructive contributions. This indicates the reliability of the
sum-rule calculation of the $f_0$ distribution amplitudes (see
Appendix B). When $\theta=0$, $f_0$ is a pure $s\bar s$ state and
hence the penguin diagram Fig. 1(a) does not contribute (i.e. the
form factor $F_0^{Bf_0^u}$ vanishes). On the other extreme with
$\theta=90^\circ$, $f_0$ is purely a $n\bar n$ state and the
penguin diagram Fig. 1(b) vanishes (i.e. $\bar f_s=0$). Since
$(a_4^p-r_\chi^K a_6^p)$ is positive, it follows from Eq.
(\ref{eq:AmpfK}) that, for a finite mixing angle, the interference
between $a_6^p(f_0K)$ and $a_6^p(Kf_0)$ penguin terms arising from
Figs. 1(a) and 1(b), respectively, is destructive for
$0<\theta<\pi/2$ and constructive for $\pi/2<\theta<\pi$. As
stated before, the $f_0\!-\!\sigma$ mixing angle is slightly
favored to be in the second quadrant, i.e. $\pi/2<\theta<\pi$. It
is evident from Fig. \ref{fig:BRqcdf} that this mixing angle
solution is also preferable by the measurement of $B\to f_0K$.
More precisely, $\B(B^-\to f_0K^-)$ is obtained to be
$(5.5-8.2)\times 10^{-6}$ for $25^\circ< \theta<40^\circ$ and
$(7.8-10.9)\times 10^{-6}$ for $140^\circ<\theta<165^\circ$.
However, even the maximal branching ratio $11.1\times 10^{-6}$
occurring at $\theta\approx 0$ is still too small by around 40\%
compared to experiment. It should be remarked that our results for
$\B(B\to f_0K)$ are larger than the previous calculations
\cite{Chen1,Chen2} owing to the large scalar decay constants
$\tilde f_{s,n}$ we have derived (see Appendix A).

\begin{figure}[t]
\vspace{0cm}
  \centerline{
            {\epsfxsize3 in \epsffile{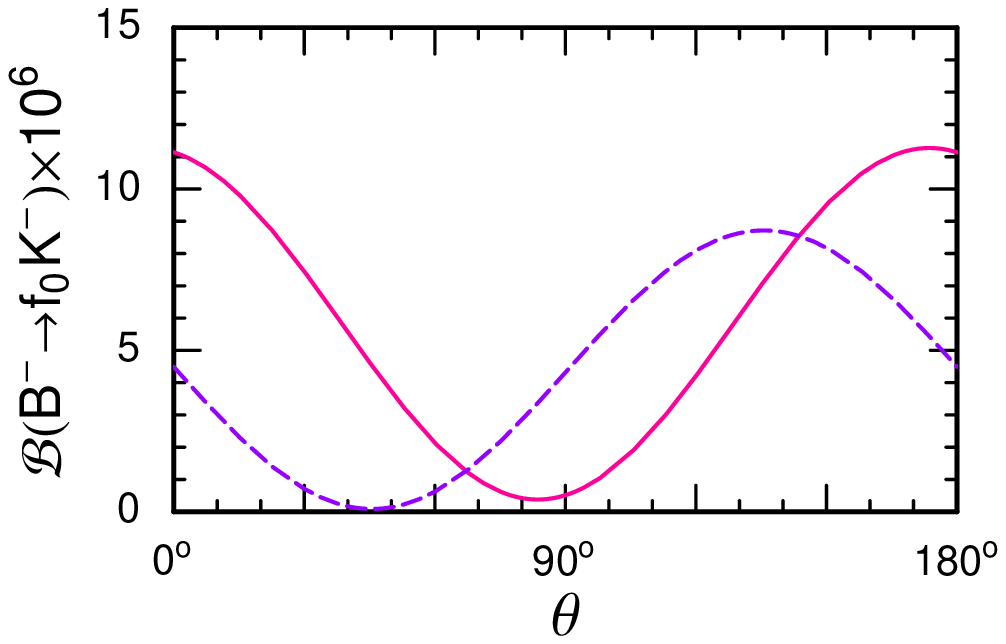}}{\epsfxsize3 in \epsffile{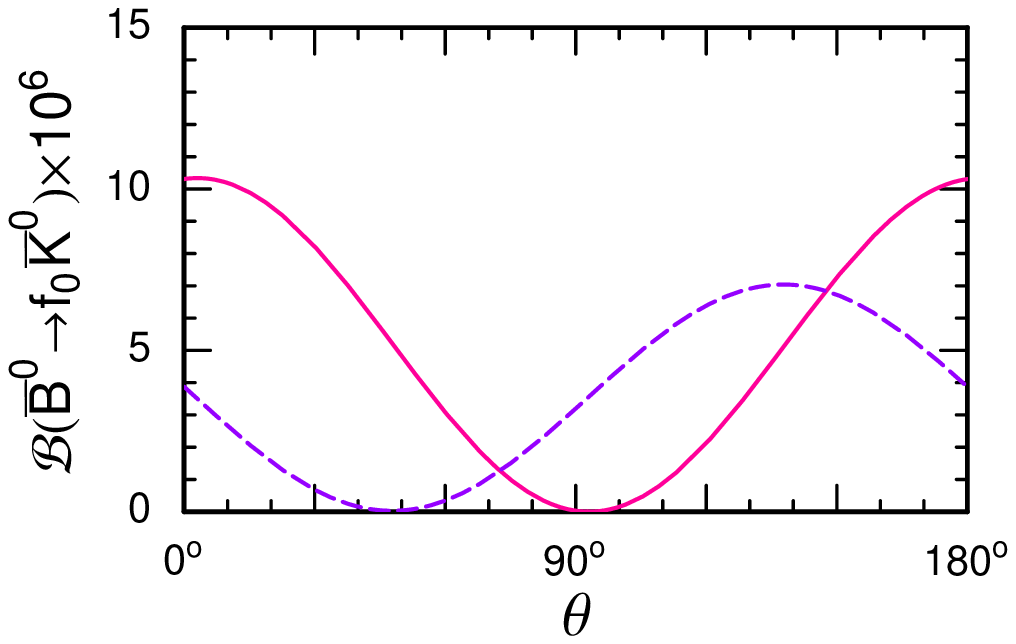}}
            }
\centerline{(a)\hspace{7.5cm} (b)\hspace{-1cm}} \vspace{0cm}
    \caption[]{\small Branching ratios of $B\to f_0(980)K$ versus the
    mixing angle $\theta$ of strange and nonstrange components of $f_0(980)$,
    where the solid curve corresponds to the Gegenbauer moments $B_{n,s}$ given in Eq.
    (\ref{app:momresults}) and the dashed curve is for the same Gegenbauer moments
    but with opposite signs. Calculations are based on QCD
    factorization.}
    \label{fig:BRqcdf}
\end{figure}

The fact that the observed $f_0(980)K$ rate is higher than the
naive model prediction calls for some mechanisms beyond the
conventional short-distance approach. Some possibilities are:

\begin{itemize}
\item Final state interactions. The predicted $B\to\pi K$ rates in
the short-distance approach are in general smaller than the data
by around 30\% (see e.g. \cite{CCS}). It is also known that the
QCDF predictions for penguin-dominated modes such as $B\to
K^*\pi,~K\phi,~K^*\phi$ are consistently lower than the data by a
factor of 2 to 3 \cite{BN}. Long-distance rescattering via charm
intermediate states (or the so-called charming penguins) will not
only enhance the aforementioned penguin-dominated decays but also
drive sizable direct $CP$ violation observed recently in $B^0\to
K^+\pi^-$ mode \cite{CCS}. The same rescattering effects may be
expected to enhance $f_0(980)K$ rates. Unfortunately, we are not
able to estimate the long-distance rescattering contributions to
the $f_0K_S$ rate from intermediate charm states owing to the
absence of information on $f_0DD$ and $f_0 D^*_{(s)}D^*_{(s)}$
couplings.

\item Gluonic coupling of the scalar meson. It is known that a
possible explanation of the enormous production of $B\to\eta' K$
and $B\to \eta'X_s$ may be ascribed to the process $b\to s+g+g$
and the two gluons fragment into $\eta'$ \cite{AS}. The same
mechanism may be also responsible for the enhancement of
$f_0(980)K$ \cite{Minkowski}.

\item Large weak annihilation contributions. Just like the pQCD
approach \cite{Keum} where the annihilation topology plays an
essential role for producing sizable strong phases and for
explaining the penguin-dominated $B\to VP$ modes, it has been
suggested in \cite{BN} (see also \cite{Kagan}) that a favorable
scenario (denoted as S4) for accommodating the observed
penguin-dominated $B\to PV$ decays and the measured sign of direct
$CP$ asymmetry in $\ov B^0\to K^-\pi^+$ is to have a large
annihilation contribution by choosing $\rho_A=1$,
$\phi_A=-55^\circ$ for $PP$, $\phi_A=-20^\circ$ for $PV$ and
$\phi_A=-70^\circ$ for $VP$ modes. The sign of $\phi_A$ is chosen
so that the direct $CP$ violation $A_{K^-\pi^+}$ agrees with the
data. Using the same set of $\rho_A$ and $\phi_A$ as the $PP$
modes, we found $\B(B^-\to f_0K^-)=17.5\times 10^{-6}$ in
agreement with the data. However, the origin of these phases is
unknown and their signs are not predicted. Moreover, since both
annihilation and hard spectator scattering encounter endpoint
divergences, there is no reason that soft gluon effects will
affect only $\rho_A$ but not $\rho_H$.

\item Subleading corrections arising from the three-parton Fock
states of the $f_0$ and/or from the intrinsic gluon inside the $B$
meson. It has been shown that the subleading corrections to QCD
factorization due to the extrinsic softer gluon effect could
enhance the branching ratio of $K\eta'$ to the level above
$50\times 10^{-6}$ ~\cite{KCYang}.
 \end{itemize}

In the present work, we have examined in Secs. II B and II C the
subleading effects mentioned in the last item. For corrections due
to the three-parton Fock states of the $f_0$ or kaon, we find that
they are of order $1/m_b^2$ and numerically negligible in $B\to
f_0 (980) K$ owing to the fact that $\la f_0(980)|V_\mu|0\ra=0$.
The detailed calculations can be found in Sec.~\ref{sec:subQCDF}
and are summarized in Eq.~(\ref{qcdfamp}). As for the corrections
originating from the intrinsic gluon within the $B$ meson, they
belong to the higher Fock components arising from quantum
fluctuations and are suppressed by $\bar\Lambda /m_b$. Since this
mechanism is nonperturbative,  the twist expansion for the $f_0$
(or $K$) distribution amplitudes is no longer suitable in this
case. Therefore, the contribution shown in Fig.~\ref{fig:igsr1}(a)
is expected to be suppressed by $1/(m_b)^n$ with $1 \leq n <2$.
Based on the reduction formula and quark-hadron duality, we are
able to estimate the nonperturbative effects in
Fig.~\ref{fig:igsr1}(a). A detailed study of intrinsic gluon
effects has been shown in Sec.~\ref{sec:intrinsicgluon} and is
summarized in Eq. (\ref{eq:totIG}).

Taking into account the aforementioned subleading corrections,  we
show in Fig.~\ref{fig:fullbr} the branching ratios of $B\to
f_0(980)K$ versus the mixing angle $\theta$ of strange and
nonstrange components of $f_0(980)$. At $\theta\sim 20^\circ$, the
intrinsic gluon effects can make 25-40\% corrections to the decay
amplitude so that the resulting branching ratio is of order
$(12\sim 20)\times 10^{-6}$, in agreement with experiment. These
corrections are constructive for $0^\circ \lesssim \theta \lesssim
80^\circ$ as well as $160^\circ \lesssim \theta \lesssim
180^\circ$ and destructive for $80^\circ \lesssim \theta \lesssim
150^\circ$. In short, when the effects due to the intrinsic
spectator gluon within the $B$ meson are inlcuded, $ 0^\circ
\lesssim \theta \lesssim 55^\circ$ and $ 155^\circ \lesssim \theta
\lesssim 180^\circ$ are preferable by the $B\to f_0 K$ data.

\begin{figure}[t]
\begin{center}
 \centerline{
            {\epsfxsize3 in \epsffile{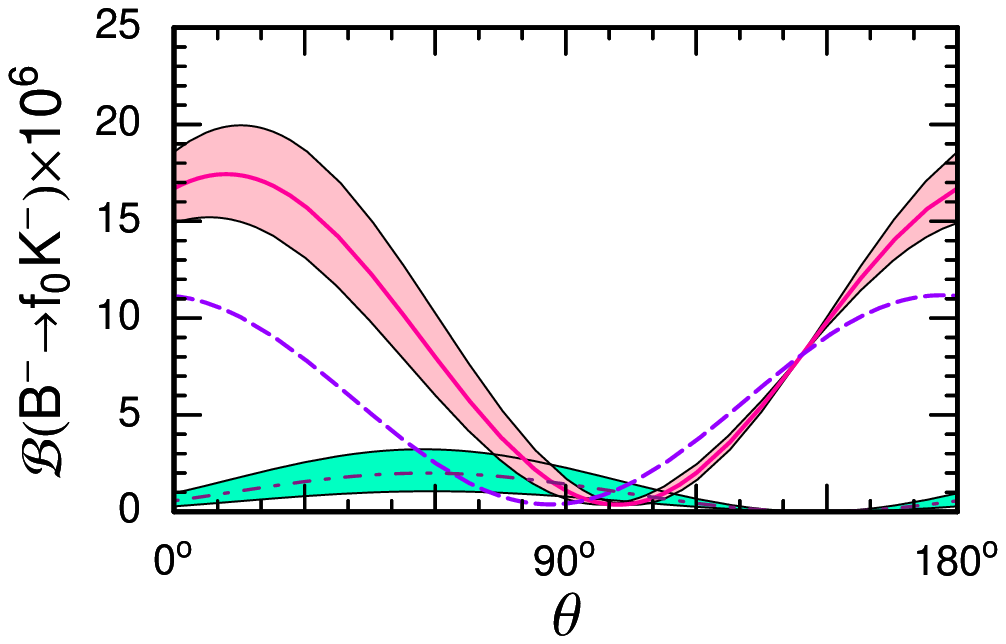}}{\epsfxsize3 in \epsffile{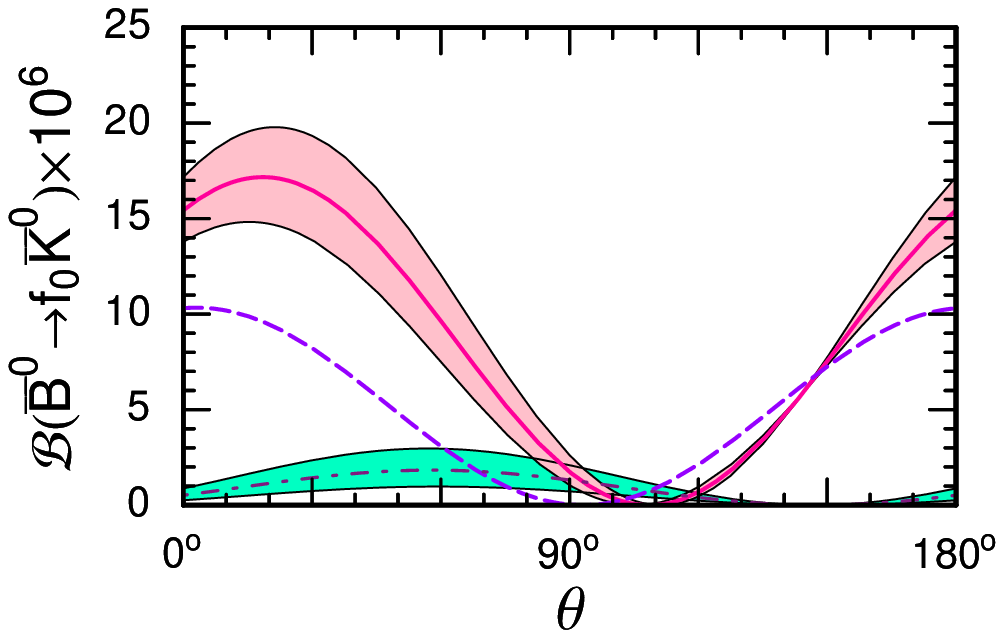}}
            }
\caption{Branching ratios of $B\to f_0(980)K$ versus the mixing
angle $\theta$ of strange and nonstrange components of $f_0(980)$,
where the solid curves are for full decay amplitudes, dashed
curves for short-distance QCD factorization amplitudes, and only
the effects from the intrinsic gluon within the $B$ meson are
evaluated in the dash-dotted curve. The bands correspond to the
errors in Eq.~(\ref{eq:o8}).} \label{fig:fullbr}
\end{center}
\end{figure}

\section{Conclusions}

In the present work we have studied the decay $B\to f_0(980)K$
within the framework of QCD factorization. Our conclusions are as
follows:

\begin{enumerate}
 \item
While it is widely believed that $f_0(980)$ is predominately a
four-quark state, in practice it is difficult to make quantitative
predictions on $B\to f_0K$ based on the four-quark picture for
$f_0(980)$ as it involves not only the $f_0$ form factors and
decays constants that are beyond the conventional quark model but
also additional nonfactorizable contributions that are difficult
to estimate. Hence, we shall assume the two-quark scenario for
$f_0(980)$.
 \item
There are two distinct penguin contributions to $B\to f_0K$ and
their interference depends on the unknown mixing angle $\theta$ of
strange and nonstrange quark contents of $f_0(980)$: destructive
for $0<\theta<\pi/2$ and constructive for $\pi/2<\theta<\pi$. A
mixing angle in the second quadrant  is preferable by the
measurement of $B\to f_0(980)K$.
 \item
Based on the QCD sum rule method, we have derived the
leading-twist light-cone distribution amplitudes of $f_0(980)$ and
the scalar decay constant $\tilde f_q$. It is found that $\tilde
f_s$ is much larger than the previous estimate owing to its scale
dependence and the large radiative corrections to the quark loops
in the OPE series. The measured $B\to f_0K$ rates clearly favor
the sign of the $f_0$ distribution amplitudes predicted by the sum
rule.
 \item
Based on the QCD factorization approach, we obtain $\B(B^-\to
f_0K^-)=(5.5-8.2)\times 10^{-6}$ for $25^\circ< \theta<40^\circ$
and $(7.8-10.9)\times 10^{-6}$ for $140^\circ<\theta<165^\circ$.
Hence, the short-distance contributions are not adequate to
explain the observed large rates of $f_0K^-$ and $f_0\ov K^0$.
 \item
Possible subleading corrections from the three-parton Fock states
of the $f_0$ and from the intrinsic (spectator) gluon inside the
$B$ meson are estimated. It is shown that while the extrinsic
gluon contribution to $B\to f_0K$ is negligible, the intrinsic
gluon within the $B$ meson may play an eminent role for the
enhancement of $f_0(980)K$.

\end{enumerate}

\vskip 2.5cm
\acknowledgments
 We are grateful to Chuang-Hung Chen
for useful discussions. This work was supported in part by the
National Science Council of R.O.C. under Grant No.
NSC93-2112-M-001-043, NSC93-2112-M-033-004.

\appendix
\section{Determination of the scalar coupling of $f_0$}
To determine the scalar decay constant $\widetilde f_q$ of
$f_0(980)$ defined by $\langle 0| \bar qq |f_0^q\rangle
=m_{f_0}^{(q)} \widetilde f_q$ with $f_0^n=\bar nn\equiv (\bar
uu+\bar dd)/\sqrt{2}$ and $f_0^s=\bar ss$, we consider the
two-point correlation function
\begin{eqnarray}\label{app:masssrf0}
\Pi (q^2)=i\int d^4x e^{iqx} \langle 0|{\rm T} (j^q(x) j^{q\dag}
(0)|0\rangle \,,
\end{eqnarray}
where $j^q=\bar qq$. The above correlation function can be
calculated from the hadron and quark-gluon dynamical points of
view, respectively. Therefore, the correlation function arising
from the lowest-lying meson $f_0^q$ can be approximately
 written as
\begin{eqnarray}
\frac{m_{f_0}^{(q)2} \widetilde f_q^2}{m_{f_0}^{(q)2}-q^2}=
\frac{1}{\pi}\int^{s_0}_0 ds \frac{{\rm Im} \Pi^{\rm OPE}}{s-q^2}
\,, \label{eq:higherresonance}
\end{eqnarray}
where $\Pi^{\rm OPE}$ is the QCD operator-product-expansion (OPE)
result at the quark-gluon level, $s_0$ is the threshold of the
higher resonant states and the contributions originating from
higher resonances are approximated by
\begin{eqnarray}
\frac{1}{\pi}\int_{s_0}^\infty ds \frac{{\rm Im} \Pi^{\rm
OPE}}{s-q^2}\,.
\end{eqnarray}
We apply the Borel transformation to both sides of
Eq.~(\ref{eq:higherresonance}) to improve the convergence of the
OPE series and suppress the contributions from higher resonances.
Consequently, the sum rule with OPE series
 up to dimension 6 and ${\cal O}(\alpha_s)$ corrections reads~\cite{GRVW}
\begin{eqnarray}
&& m_{f_0}^{(q)2} \widetilde f_{q}^2 e^{-m_{f_0}^{(q)2}/M^2}\Bigg
(\frac{\alpha_s(\mu)}{\alpha_s(M)}\Bigg)^{8/b} = \frac{3}{8\pi^2}
M^4 \Biggl[ 1 + \frac{\alpha_s(M)}{\pi}\Bigg( \frac{17}{3} +2
\frac{I(1)}{f(1)}-2\ln \frac{M^2}{\mu^2} \Bigg)f(1) \Biggr]
\nonumber\\
&&\ \ + \frac{1}{8}\langle \frac{\alpha_s
G^2}{\pi}\rangle+3\langle m_q \bar qq\rangle
-\frac{\pi\alpha_s}{M^2}\Bigg(\langle\bar q\sigma_{\mu\nu}T^a q \,
\bar q\sigma^{\mu\nu}T^a q\rangle + \frac{2}{3}\langle\bar
q\gamma_\mu T^a q \, \bar q\gamma^\mu T^a q\rangle \Bigg) \,,
\label{eq:appf0masssr}
\end{eqnarray}
where $f(1)=1-e^{-s_0/M^2}(1+s_0/M^2)$,
$I(1)=\int^1_{e^{-s_0/M^2}}(\ln t) \ln (-\ln t)dt$, and we have
taken into account the scale dependence of  $\widetilde f_q$,
\begin{eqnarray}
\widetilde f_q (M)=\widetilde f_q(\mu)\Bigg(
\frac{\alpha_s(\mu)}{\alpha_s(M)} \Bigg)^{4/b},
\end{eqnarray}
with $b=(11 N_c -2n_f)/3$. Here, the anomalous dimensions of
$\alpha_s G^2$ and  $m_q \bar qq$ are equal to zero, while the
anomalous dimensions of the 4-quark operators have been neglected.
In the numerical analysis, we shall use the following values for
vacuum condensates and quark masses at the scale of
$\mu=1$~GeV~\cite{Yang:1993bp}:
\begin{eqnarray}
\begin{array}{lcl}
  \langle \alpha_s G_{\mu\nu}^a G^{a\mu\nu} \rangle=0.474\ {\rm GeV}^4/(4\pi)\,, &  &   \\
  \langle \bar uu \rangle \cong \langle \bar dd \rangle =-(0.24\ {\rm
  GeV})^3 \,,
  &\ \ \  & \langle \bar ss \rangle = 0.8 \langle \bar uu \rangle \,, \\
  (m_u+m_d)/2=5\ {\rm MeV}\,, &  & m_s=119\ {\rm MeV}\,,
\end{array}\label{eq:parameters}
\end{eqnarray}
and adopt the vacuum saturation approximation for describing the
four-quark condensates, i.e.,
\begin{eqnarray}
\langle 0|\bar q \Gamma_i T^a q \bar q \Gamma_i T^a q|0\rangle
=-\frac{1}{16N_c^2}{\rm Tr}(\Gamma_i\Gamma_i) {\rm Tr}(T^a T^a)
\langle \bar qq\rangle^2 \,.
\end{eqnarray}
Taking the logarithm of both sides of Eq.~(\ref{eq:appf0masssr})
and then applying the differential operator $M^4 \partial
/\partial M^2$ to them, we obtain the mass sum rule for
$f_0^{(q)}$. In Fig.~\ref{fig:massf0}, we explore two scenarios
for (i) $\bar qq= \bar ss$ and (ii) $\bar qq=\bar nn$.
Numerically, we get
\begin{eqnarray}
&& m_{f_0}^{(s)}\simeq (1.02\pm 0.05)~{\rm GeV} \ \ \ \
 {\rm for\ s_0=2.6\ GeV^2}\,, \nonumber\\
&& m_{f_0}^{(n)}\simeq (0.99\pm 0.05)~{\rm GeV} \ \ \ \
 {\rm for\ s_0=2.6\ GeV^2}\,,
\end{eqnarray}
where the value of $s_0$ is determined when the maximum stability
for the mass sum rule is reached. Substituting the above results
for the threshold $s_0$ and the masses obtained in the mass sum
rule into Eq.~(\ref{eq:appf0masssr}), we obtain the sum rule for
$\widetilde f_q$ at $\mu=1$~GeV as a function of the Borel mass
squared $M^2$ (see Fig.~\ref{fig:massf0}). The results are
\begin{eqnarray} \label{eq:fs1}
 \widetilde f_s(\mu=1~{\rm GeV})= 0.33~{\rm GeV}\,, \ \ \ \
\  \widetilde f_n(\mu=1~{\rm GeV})\simeq 0.35~{\rm GeV}\,,
\end{eqnarray}
and
\begin{eqnarray} \label{eq:fs2.1}
 \widetilde f_s(\mu=2.1~{\rm GeV})\simeq 0.39~{\rm GeV}\,, \ \ \ \
\  \widetilde f_n(\mu=2.1~{\rm GeV})\simeq 0.41~{\rm GeV}\,.
\end{eqnarray}
Evidently, they are much larger than the typical decay constant of
pseudoscalar mesons.

Two remarks are in order. First, the scale dependence of
$\widetilde f_q$ was not considered in \cite{Fazio,Bediaga}. If
the scale dependence is neglected, we will obtain $s_0\simeq
1.6~{\rm GeV}^2$, in accordance with the results in
\cite{Fazio,Bediaga}. Second, the large radiative correction to
the quark loop in the OPE series arises mainly from one-gluon
exchange and is likely the effect of the color Coulomb
interaction, as in the case for the $B$ meson~\cite{Bagan:1991sg}.
The values of $\widetilde f_q$ and $s_0$, rather than
$m_{f_0}^{(q)}$,  are very sensitive to the radiative corrections
to the quark loop and to the scale dependence of $\widetilde f_q$.
However, these effects were not considered in
\cite{Fazio,Bediaga}. If neglecting $\alpha_s$ corrections to the
mass sum rule while keeping $s_0=2.6$~GeV$^2$, the resulting
$\widetilde f_s(\mu$=1~GeV) will be 0.26~GeV. Moreover, if further
setting $s_0=1.6$~GeV$^2$, the sum rule yields $\widetilde
f_s(\mu$=1~GeV)=0.19~GeV, in agreement with \cite{Fazio,Bediaga}.
The validity of our results is thus realized. It is interesting to
note that a larger scalar decay constant for $K_0^*(1430)$ has
been obtained in \cite{Chernyak,Du} based on the sum-rule method.
\begin{figure}[t]
\vspace{0cm}
\centerline{\psfig{figure=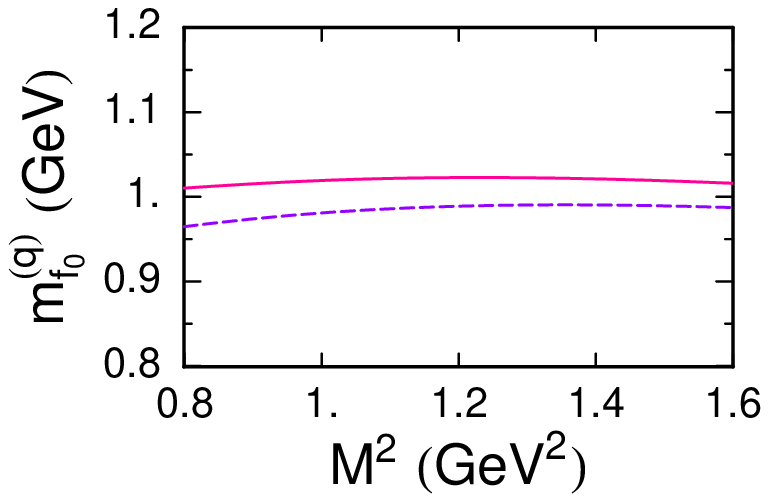,width=6cm}\hskip1cm
\psfig{figure=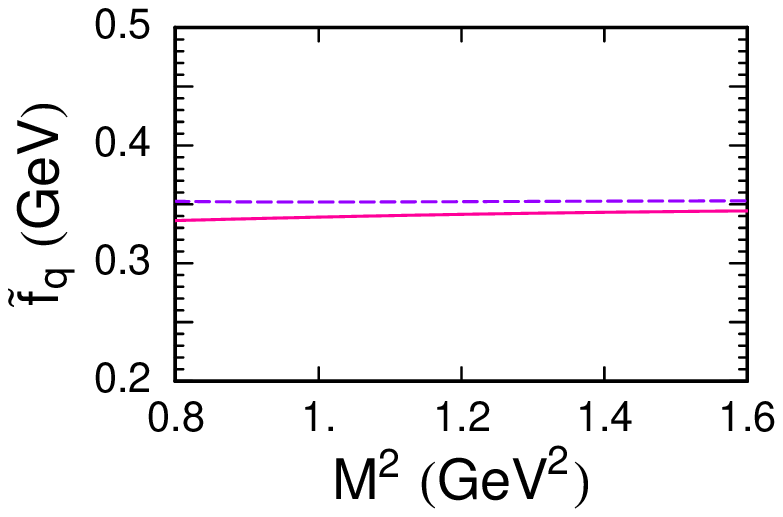,width=6cm} } \vspace{0cm} \caption[]{\small
$m_{f_0}^{(q)}$ and $\widetilde f_{q}$ as functions of the Borel
mass squared $M^2$. The solid curve is obtained for $j^s=\bar s s$
and the dashed curve for $j^n=\bar n n$. } \label{fig:massf0}
\end{figure}

\section{Leading twist LCDA for $f_0$}
The LCDAs $\Phi_{q}(x,\mu)$ corresponding to the ideal states
$f^q_0=\bar qq$ are defined by
 \be
 \la f_0^q(p)|\bar q(z)\gamma_\mu q(0)|0\ra &=& p_\mu \tilde f_q\int
 ^1_0 dx e^{ixp\cdot z}\Phi_{q}(x,\mu),
 \en
where $x$ (or $\bar x=1-x$) is the $f_0$ momentum fraction carried
by the quark $q$ (or antiquark $\bar q$) and $\mu$ is the
normalization scale of the LCDA. $\Phi_{q}(x,\mu)$ can be expanded
in a series of Gegenbauer polynomials~\cite{Chernyak:1983ej,Braun}
\begin{eqnarray}\label{app:generalda}
\Phi_{q}(x,\mu)=6x(1-x)\sum_{l=1,3,5,\dots} \phi_l^{q}(\mu)
C^{3/2}_l(2x-1),
\end{eqnarray}
with multiplicatively renormalizable coefficients (or the
so-called Gegenbauer moments):
\begin{eqnarray}
\phi_l^{q}(\mu) = \frac{2(2l+3)}{3(l+1)(l+2)} \int_0^1 C^{3/2}_l
(2x-1) \Phi_{q}(x,\mu).
\end{eqnarray}
Consider the following two-point correlation function
\begin{eqnarray}
\Pi_l (q) = i\int d^4x e^{iqx} \langle 0| T(O_l(x)\ O^\dagger(0)
|0 \rangle = (zq)^{l+1} I_l (q^2),
\end{eqnarray}
where
\begin{eqnarray}
&&\langle 0| O_l|f_0^q(p)\rangle \equiv \langle 0| \bar q \not\! z
(i z \stackrel{\leftrightarrow}{D})^l q |f_0^q(p)\rangle=(zp)
\widetilde{f}_q  \int^1_0 (2x-1)^2
\Phi_{q}(x) dx \equiv (zp) \widetilde{f}_q \langle \xi^l_{q} \rangle,\\
&& \langle 0| O|f_0^q(p)\rangle \equiv \langle 0| \bar q q
|f_0^q(p)\rangle = m^{(q)}_{f_0} \widetilde{f}_q,
\end{eqnarray}
with $z^2=0$. The Gegenbauer moments $\phi_l^{q}(x,\mu)$ can be
easily expressed in terms of the above defined moments $\langle
\xi^l_q \rangle$, for which the sum rule reads
\begin{eqnarray}\label{app:srmoments}
\langle \xi^l_{q}\rangle = \frac{1}{m^{(q)}_{f_0}
\widetilde{f}_q^2} e^{m^{(q)2}_{f_0} /M^2} [(-1)^{l+1} +1
]\Bigg(\langle \bar qq\rangle + \frac{3l-1}{24} \frac{\langle \bar
q g_s \sigma\cdot G q \rangle}{M^2} + \frac{l(l-1)}{48}{\langle
g_s^2 G^2\rangle \langle \bar qq\rangle \over M^4} \Bigg).
 \end{eqnarray}
The conformal invariance in QCD exhibits that partial waves, in
the expansion of $\Phi_{q}(x,\mu)$ in Eq.~(\ref{app:generalda}),
with different conformal spin cannot mix under renormalization to
the leading-order accuracy. As a consequence, the Gegenbauer
moments $\phi_l$ in (\ref{app:generalda}) renormalize
multiplicatively:
  \begin{equation}
    \phi_l(\mu) = \phi_l(\mu_0)
  \left(\frac{\alpha_s(\mu_0)}{\alpha_s(\mu)}\right)^{-\gamma_{(l)}/{b}},
  \label{momentdar}
   \end{equation}
where the one-loop anomalous dimensions are \cite{GW}
  \begin{eqnarray}
  \gamma_{(l)}  = C_F
  \left(1-\frac{2}{(l+1)(l+2)}+4 \sum_{j=2}^{l+1} \frac{1}{j}\right),
  \label{eq:1loopandim}
  \end{eqnarray}
with $C_F=(N_c^2-1)/(2N_c )$.  We consider the
renormalization-improved sum rule of Eq.~(\ref{app:srmoments}),
where the anomalous dimensions of relevant operators can be found
in~\cite{Yang:1993bp}, such that
 \begin{eqnarray}\label{app:anamolous}
 &&\langle \bar q q\rangle_\mu = \langle \bar q q\rangle_{\mu_0}
 \left(\frac{\alpha_s(\mu_0)}{\alpha_s(\mu)}\right)^{4\over b},\nonumber\\
 && \langle g_s \bar q\sigma\cdot G q\rangle_\mu =
 \langle g_s \bar q\sigma\cdot G q\rangle_{\mu_0}
 \left(\frac{\alpha_s(\mu_0)}{\alpha_s(\mu)}\right)^{-{2\over 3b}},\nonumber\\
 && \langle g_s^2 G^2\rangle_\mu = \langle g_s^2 G^2\rangle_{\mu_0} .
 \end{eqnarray}

In the numerical analysis, we choose the Borel window
$0.9$~GeV$^2< M^2 < 1.2$~GeV$^2$, consistent with the previous
case, where the contributions originating from higher resonances
and the highest OPE terms are well under control. We also use the
input parameters given in Eq.~(\ref{eq:parameters})
and~\cite{Yang:1993bp}
\begin{eqnarray}\label{app:mixcondensate}
&& \langle g_s \bar u\sigma Gu \rangle \cong \langle g_s\bar
d\sigma Gd \rangle =-0.8\langle \bar uu \rangle,\nonumber\\
&& \langle g_s \bar s\sigma Gs \rangle = 0.8 \langle g_s\bar
u\sigma Gu \rangle.
 \end{eqnarray}
It should be noted that in the large $l$ limit for the moment
$\langle\xi^l\rangle$ sum rule, the actual expansion parameter is
$M^2/l$. As a result, for $l\geq 7$ and fixed $M^2$, the OPE
series are convergent slowly or even divergent, i.e. the resulting
sum-rule result becomes less reliable.
 We thus obtain the first three non-zero moments, at the normalization
 scale $\mu=1$~GeV,
 \begin{eqnarray}\label{app:momresults}
&&  \langle \xi_{n}^1 \rangle = - 0.55\pm 0.05\ \Rightarrow\
 B_1^{(n)}\equiv\phi_1^n= 5\langle \xi_{n}^1\rangle/3 = - 0.92\pm 0.08,\nonumber\\
&& \langle \xi^3_{n} \rangle = - 0.43\pm 0.02\ \Rightarrow\
 B_3^{(n)}\equiv\phi_3^n= \frac{21}{4}\langle \xi^3_{n}\rangle
 -\frac{9}{4}\langle \xi_{n}^1 \rangle = - 1.00 \pm
 0.05,\nonumber\\
 &&  \langle \xi_{n}^5 \rangle = - 0.33\pm 0.01\ \Rightarrow\
 B_5^{(n)}\equiv\phi_5^n= \frac{3}{4}(7\langle \xi_{n}^3\rangle-3\langle
 \xi_{n}^1\rangle)
 = - 0.40\pm 0.05,
 \end{eqnarray}
where the sub-(super-)script $n$ denotes $\bar nn$. Rescaling to
the normalization point $\mu=2.1$~GeV, we find
 \begin{eqnarray}
&&  \langle \xi_{n}^1 \rangle = - 0.44\pm 0.04\ \Rightarrow\
 B_1^{(n)}\equiv\phi_1^n= - 0.73\pm 0.07,\nonumber\\
&& \langle \xi^3_{n} \rangle = - 0.33\pm 0.02\ \Rightarrow\
 B_3^{(n)}\equiv\phi_3^n = - 0.74 \pm 0.04, \nonumber\\
 &&  \langle \xi_{n}^5 \rangle = - 0.24\pm 0.01\ \Rightarrow\
 B_5^{(n)}\equiv\phi_5^n = - 0.017\pm 0.005.
 \end{eqnarray}
As for $\bar qq=\bar ss$, the analysis yields
$\langle\xi_{s}^{1,3,5} \rangle \simeq  (\langle \bar ss\rangle
/\langle \bar uu\rangle) \langle\xi_{n}^{1,3,5} \rangle \simeq 0.8
\langle\xi_{n}^{1,3,5} \rangle$ and $B_{1,3,5}^{(s)} \simeq 0.8
B_{1,3,5}^{(n)}$. The leading-twist LCDAs $\Phi_{s}(x,\mu)$ and
$\Phi_{n}(x,\mu)$ for the $f_0$ meson, which vanish in the large
$\mu$ limit,  are illustrated in Fig.~\ref{fig:das}.

\begin{figure}[t]
\vspace{0cm}
\centerline{\psfig{figure=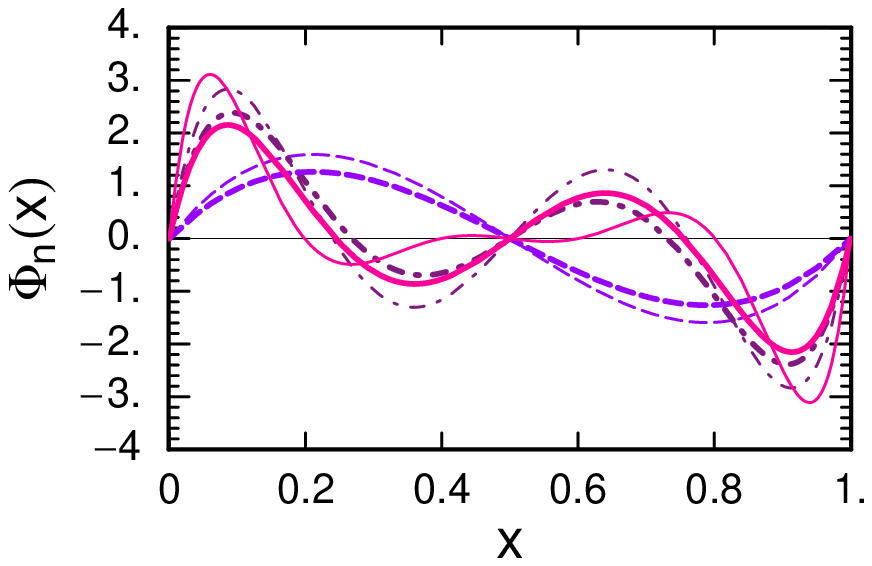,width=6cm}\hskip1cm
\psfig{figure=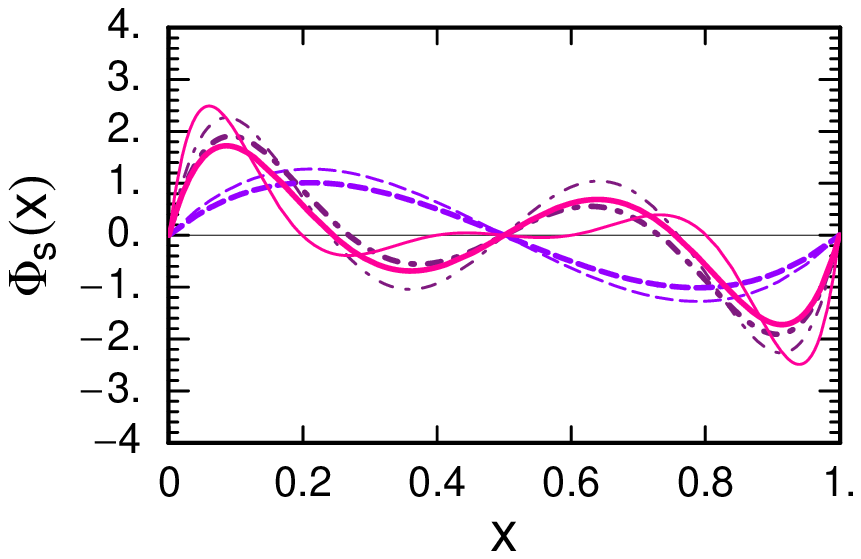,width=6cm} } \vspace{0cm} \caption[]{\small
The leading twist-2 LCDAs for the $f_0$ meson vs the momentum
fraction carried by the quark, where the solid curves are
evaluated with considering the $B_1, B_3, B_5$ central values
given in Eq.~(\ref{app:momresults}), while the dash-dotted and
dashed lines are obtained by setting $B_5=0$ and $B_3=B_5=0$,
respectively. The darker lines correspond to the normalization
scale $\mu=2.1$~GeV, while the lighter lines to $\mu=1$~GeV.}
\label{fig:das}
\end{figure}

\section{ Decay constant $\delta_{f^q_0}^2$}\label{app:c}
The parameter  $\delta_{f^q_0}^2$ appearing in
Eq.~(\ref{eq:3pdas}) can be defined through the matrix element of
the local current operator $\widetilde j_\beta^q=\bar q
\gamma^\alpha g_s G_{\alpha\beta} q$,
\begin{eqnarray}
\langle 0|\widetilde j^q_\beta(0) |f_0^q(p)\rangle = i \widetilde
f_q p_\beta \delta^2_{f_{0}^q}\,.
\end{eqnarray}
To calculate the $\delta_{f_0}$ parameter, we consider the
following non-diagonal two-point correlation function
\begin{eqnarray}
\widetilde \Pi_\beta (q^2)=i\int d^4x e^{iqx} \langle 0|{\rm T}
(\widetilde j^q_\beta (x) j^{q\dag} (0))|0\rangle\ =q_\beta
\widetilde \Pi(q^2),
\end{eqnarray}
where $j^q=\bar qq$ as in Eq.~(\ref{app:masssrf0}). In the deep
Euclidean region of $q^2$, the above correlation function can be
perturbatively calculated in QCD and the result is given by
 \begin{eqnarray}
\widetilde \Pi^{\rm QCD} (q^2) = -i\frac{2\alpha_s}{3\pi}\langle
\bar qq\rangle \Bigg(\ln (-q^2/\mu^2)+ \frac{7}{6}\Bigg)  -
\frac{i}{2q^2} \langle \bar q g_s \sigma\cdot Gq\rangle \,.
 \end{eqnarray}
After approximating the higher resonances as given in
Eq.~(\ref{eq:higherresonance}) and performing  the Borel
transformation, the resultant sum rule with OPE series up to the
order of dimension 7 and radiative corrections of order ${\cal
O}(\alpha_s)$ reads
\begin{eqnarray}\label{eq:f0delta}
&& e^{-m^{(q)}_{f_0}/M^2} \widetilde f_q^2 \delta_{f_0^q}^2
m^{(q)}_{f_0}\Bigg({\alpha_s(\mu)\over
\alpha_s(M)}\Bigg)^{-{32\over 9b}} \Bigg({\alpha_s(\mu)\over
\alpha_s(M)}\Bigg)^{{4\over b}}\nonumber\\
&& = \frac{2}{3} \frac{\alpha_s}{\pi}\langle\bar qq \rangle
\Bigg({\alpha_s(\mu)\over \alpha_s(M)}\Bigg)^{4\over b} M^2
(1-e^{-s_0/M^2}) + \frac{1}{2}\langle \bar q g_s \sigma\cdot
Gq\rangle \Bigg({\alpha_s(\mu)\over \alpha_s(M)}\Bigg)^{-{2\over
3b}} \,,
\end{eqnarray}
where the anomalous dimensions of some operators have been listed
in Eq.~(\ref{app:anamolous}) and that of $\bar q \gamma^\alpha g_s
G_{\alpha\beta} q$ is $-32/(9b)$~\cite{Chetyrkin:2000tj}.

 In the Borel window $0.9\ {\rm GeV}^2 < M^2
< 1.2\ {\rm GeV}^2$ where the contributions originating from
higher resonances and the highest OPE term are well under control,
using input parameters in Eqs.~(\ref{eq:parameters}) and
(\ref{app:mixcondensate}), we obtain
\begin{eqnarray}
&& \delta_{f_0^s}^2(\mu=1~{\rm GeV}) = (0.09\pm 0.01)~{\rm
GeV}^2\,,
\nonumber\\
&& \delta_{f_0^n}^2(\mu=1~{\rm GeV}) = (0.10\pm 0.01)~{\rm
GeV}^2\,,
\end{eqnarray}
and, as rescaled to $\mu=2.1$~GeV,
\begin{eqnarray}
&& \delta_{f_0^s}^2(\mu=2.1~{\rm GeV}) = (0.08\pm 0.01)~{\rm
GeV}^2\,,
\nonumber\\
&& \delta_{f_0^n}^2(\mu=2.1~{\rm GeV}) = (0.09\pm 0.01)~{\rm
GeV}^2\,,\label{app:parameter3f2.1}
\end{eqnarray}
 for $j^q= \bar ss$ and $\bar nn$. The results are depicted in
Fig.~\ref{fig:f0delta}.
\begin{figure}[t]
\vspace{0cm}
 \centerline{\psfig{figure=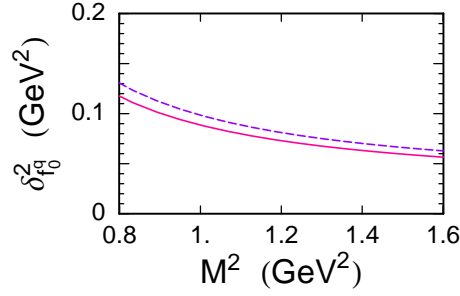,width=6cm}}
\vspace{0cm} \caption[]{\small $\delta_{f_0^q}^2(\mu=1$~GeV) as a
function of the Borel mass squared $M^2$ with the same notation as
Fig.~\ref{fig:massf0}. We have used $m_{f_0}^{(s)}=1.02$~GeV and
$\widetilde f_{s}=0.33~{\rm GeV}$ for the solid curve, while
$m_{f_0}^{(n)}=0.98$~GeV and $\widetilde f_{n}=0.35~{\rm GeV}$ for
the dashed curve.}\label{fig:f0delta}
\end{figure}

\section{Determination of $\delta_B^2$}
The parameter related the intrinsic gluon content of the $B$ meson
is defined by
\begin{eqnarray}
\langle 0| \bar u \gamma^\alpha g_s \widetilde G_{\alpha\mu}
b|B\rangle = if_B\delta_B^2 p_\mu.
\end{eqnarray}
To estimate the parameter $\delta_B^2$, we  consider the
non-diagonal two-point propagator
\begin{eqnarray}
\Pi^B_\mu=i \int d^4 x e^{iqx} \langle 0| T( \bar b(x) g_s
\widetilde G_{\alpha\mu}(x) \gamma^\alpha u(x)\ \bar u(0)\gamma_5
b(0) |0\rangle.
\end{eqnarray}
Following the same line as in Appendix~\ref{app:c}, we arrive at
the $\delta_B^2$ sum rule
\begin{eqnarray}
\delta_B^2\simeq \frac{m_b +m_u}{4 m_B^2 f_B^2}
e^{{m_B^2-m_b^2}\over M^2} \langle \bar q g_s \sigma\cdot G
q\rangle,
\end{eqnarray}
where the contribution arising from  $\alpha_s\langle \bar q
q\rangle $ is much smaller than the quark-gluon mixed condensate
shown in Eq.~(\ref{eq:f0delta}) and hence can be neglected. Using
the value for quark-gluon mixing condensate in
Eq.~(\ref{app:mixcondensate}), $f_B=185$~MeV, and $m_b=4.85$~GeV
for the pole mass of the $b$ quark, we obtain
\begin{eqnarray}
\delta_B^2 = 0.022 \pm 0.002\ {\rm
GeV}^2,\label{app:parameterintrinsic}
\end{eqnarray}
within the Borel window 9 GeV$^2 < M^2 <
12$~GeV$^2$~\cite{Yang:1997nh}. Note that the scale dependence of
$\delta^2_B$ is weak.


\end{document}